\documentclass[%
reprint,
superscriptaddress,
 amsmath,amssymb,
 aps,
showkeys
]{revtex4-2}

\usepackage{graphicx}
\usepackage{dcolumn}
\usepackage{bm}
\usepackage{xcolor}

\usepackage[colorlinks=true, citecolor=black, urlcolor=blue, linkcolor=blue, filecolor=blue]{hyperref}

\usepackage{bibunits}
\defaultbibliographystyle{apsrev4-2}

\usepackage{lipsum}
\newcommand{\savg}{\langle s \rangle}
\newcommand{\nclus}{n_{\mathrm{cluster}}}

\makeatletter
\def\selectlanguage#1{}

\makeatother

\begin{document}

\begin{bibunit}

\title{Shape and friction govern clogging through local rigidity}

\author{Jules Tampier}
\affiliation{Laboratoire de Physique et Mécanique des Milieux Hétérogènes (PMMH), UMR 7636 CNRS - ESPCI Paris - Sorbonne Université – Université Paris Cité, France}

\author{Philippe Bourrianne}
\affiliation{Laboratoire de Physique et Mécanique des Milieux Hétérogènes (PMMH), UMR 7636 CNRS - ESPCI Paris - Sorbonne Université – Université Paris Cité, France}

\author{Anke Lindner}
\affiliation{Laboratoire de Physique et Mécanique des Milieux Hétérogènes (PMMH), UMR 7636 CNRS - ESPCI Paris - Sorbonne Université – Université Paris Cité, France}
\affiliation{Institut Universitaire de France (IUF), France}

\author{David Richard}
\affiliation{Laboratoire de Physique et Mécanique des Milieux Hétérogènes (PMMH), UMR 7636 CNRS - ESPCI Paris - Sorbonne Université – Université Paris Cité, France}

\date{\today}

\begin{abstract}
\textbf{
Granular flows in confined geometries can abruptly arrest due to the formation of a spanning arch or dome of particles. While the likelihood of such clogging events has been described empirically for over two decades, the physical mechanism governing clog formation and its dependence on shape remain elusive.
Here we show that the probability of flow arrest is fundamentally controlled by the number of particles required to form rigid spanning structures during discharge, which is set by a balance between degrees of freedom and constraints that depends on shape and friction.
Combining 2D microfluidic experiments and discrete numerical simulations with a combinatorial algorithm for rigid cluster decomposition, we find that the minimum number of particles required for clogging scales quadratically with the arch size when grains are free to slide or roll, and linearly when both sliding and rolling are blocked.
These results identify rigidity as the organizing principle behind granular clogging, providing a mechanistic foundation for a long-standing empirical law and a general framework that links microscopic grain mobility to macroscopic flow arrest.
}
\end{abstract}

\maketitle

Driven particulate systems display complex collective dynamics, where simple interactions between particles lead to a rich variety of phenomena~\cite{jaeger_granular_1996,goyon_spatial_2008,gu_emergence_2025}. Among these is the ability of amorphous systems such as gels~\cite{bantawa_hidden_2023}, biological tissues~\cite{bi_density-independent_2015}, and granular materials~\cite{liu_jamming_1998,dashti_emergence_2023} to transition abruptly from a flowing, fluid-like state to a rigid, solid-like state. Such a transition can typically be observed during granular flows in confined geometries, where the formation of a spanning arch (in 2D) or dome (in 3D) can suddenly interrupt the discharge of particles. This phenomenon, known as clogging \cite{zuriguel_clogging_2014,dressaire_clogging_2016,marin_clogging_2025}, poses fundamental challenges in both dry systems and suspension flows, spanning contexts from industrial bulk solid handling~\cite{schulze_powders_2021} and microfluidics~\cite{bakhtiari_method_2024} to blood~\cite{shelby_microfluidic_2003} and debris flows~\cite{piton_debris_2022}.

Clogging is commonly modeled as a stochastic process in which each exiting particle has the same probability $p$ of triggering a clog \cite{clement_jamming_2000,zuriguel_jamming_2005}. In practice, this probability is quantified by measuring the average avalanche size, $\savg = (1-p)/p$, defined as the mean number of particles escaping before clog formation \cite{janda_jamming_2008}. Although $\savg$ depends on a wide range of parameters, including particle properties \cite{hong_clogging_2017,hsu_roughness-dependent_2021,vani_influence_2022,hoggarth_clogging_2026} and channel geometry \cite{zuriguel_silo_2011,lopez-rodriguez_effect_2019,vani_role_2024}, it is primarily governed by the ratio between the outlet width $w$ and the particle size $d$ \cite{to_jamming_2005,janda_jamming_2008}. For circular or spherical particles, this dependence is generally fitted using an exponential scaling:
\begin{equation}
\ln(\savg + 1) \sim (w/d)^\alpha,
\label{eq:scaling}
\end{equation}
where $\alpha$ matches the dimensionality of the system~\cite{to_jamming_2005,janda_jamming_2008,thomas_fraction_2015,montero_dimensional_2025}.

Despite over two decades of extensive use, this scaling remains purely empirical and lacks a rigorous theoretical foundation. While some studies relate $\alpha$ to the existence of a stabilizing region of thickness $\sim w$ above the constriction  \cite{thomas_fraction_2015,hoggarth_clogging_2026}, this assumption has never been experimentally or numerically verified. Furthermore, results for polygonal and rod-like particles reveal a breakdown of the scaling, with $\savg$ increasing more slowly than predicted by equation~(\ref{eq:scaling}) \cite{goldberg_clogging_2018,ashour_outflow_2017}. This discrepancy indicates that the widely applied exponential scaling is not universal, highlighting the need for a broader physics-based framework that accounts for particle shape and other grain properties. To date, no such framework exists, leaving the underlying mechanism governing clog formation an open question.

In this Article, we establish that the probability of granular clogging is fundamentally controlled by the number of particles required to form rigid spanning structures during discharge. By combining microfluidic experiments, discrete numerical simulations, and the pebble game algorithm for rigid cluster decomposition, we demonstrate that this number is dictated by the ability of particles to slide and roll relative to each other, which ultimately depends on shape and friction. Specifically, we show that the number of particles involved in clog stability scales quadratically with the arch size when particles are free to slide or roll, while it scales linearly when both sliding and rolling are blocked. These results resolve a long-standing problem by uncovering the physical mechanism governing clog formation in confined geometries and its dependence on complex particle properties.

\begin{figure*}
\includegraphics[width=1\linewidth]{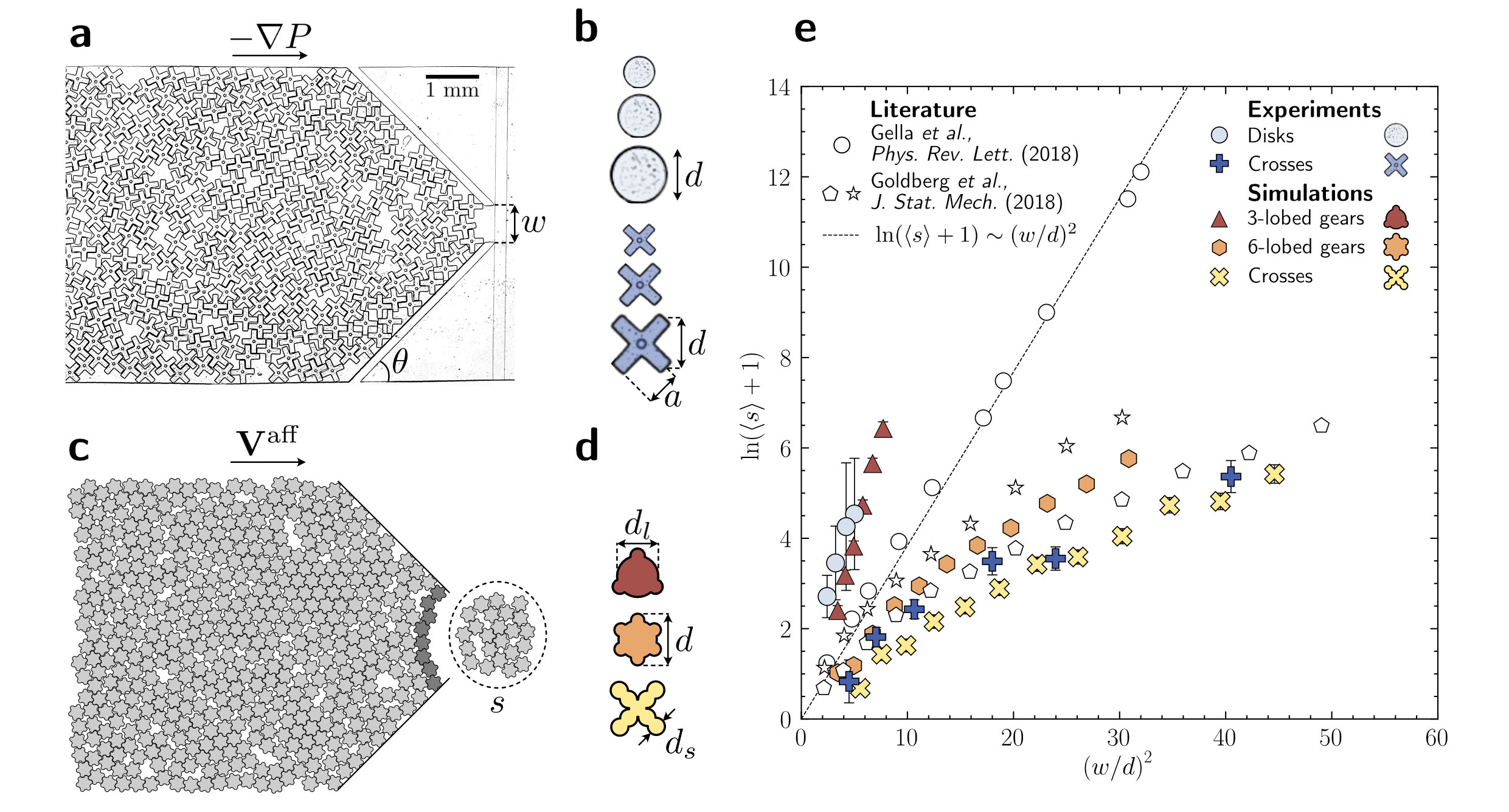}
\caption{\label{fig:fig1} \textbf{Influence of shape on the average avalanche size.} \textbf{(a)} 2D microfluidic hopper channel consisting of a straight section followed by a tapered section with angle $\theta=45^{\circ}$ and constriction width $w$. The channel is filled with \textbf{(b)} disks of diameter $d$ and crosses with arm length $a$ and characteristic size $d=\sqrt{2}a$, driven towards the constriction by a pressure gradient $\nabla P$. \textbf{(c)} Non-smooth contact dynamics simulations where non-circular particles are driven at a constant velocity $V^{\mathrm{aff}}$ through the same hopper geometry as in experiments. The simulated shapes include \textbf{(d)} gears and crosses made of rigidly connected hard disks of diameters $d_s$ and $d_l$, with characteristic sizes $d= d_s + d_l$ and $d=2\sqrt{2}d_s$, respectively. \textbf{(e)} Avalanche size $\ln(\savg + 1)$ versus the squared size ratio $(w/d)^2$. Data include previous experiments with frictional circular particles (white circles) \cite{gella_decoupling_2018} and simulations with frictional squares (white squares) and pentagons (white pentagons) \cite{goldberg_clogging_2018}. These are shown alongside our microfluidic experiments with disks (light blue circles) and crosses (dark blue crosses), and our numerical simulations with gears (red triangles, orange hexagons) and crosses (yellow crosses). The dashed line represents the fit of equation~(\ref{eq:scaling}) to the literature data for circular particles. Error bars show the standard error assuming a binomial proportion with an estimated clogging probability $p = 1 / (\savg + 1)$.}
\end{figure*}

To verify the shape-dependent deviation from equation~(\ref{eq:scaling}), we first perform clogging experiments in a quasi-2D microfluidic hopper, where \textit{in situ} fabricated particles of tunable shape are driven by a pressure gradient $\mathbf{\nabla} P$ through a tapered section with angle $\theta=45^{\circ}$ and aperture width $w$ (Fig.~\ref{fig:fig1}(a)) \cite{kool_hopper_2025}. Specifically, we compare the clogging statistics of disks of diameter $d$ and crosses with arm length $a$ and characteristic size $d=\sqrt{2}a$ (Fig.~\ref{fig:fig1}(b); Supplementary Videos~\textcolor{blue}{S1} and \textcolor{blue}{S2}).
The evolution of $\ln(\savg+1)$ with the squared size ratio for both shapes is shown in Fig.~\ref{fig:fig1}(e), alongside previous results from the literature. While the data for disks is compatible with a linear increase of $\ln(\savg+1)$ with $(w/d)^2$, crosses exhibit a markedly weaker dependence on the size ratio, as previously observed in numerical simulations with faceted particles \cite{goldberg_clogging_2018}.

To explore a broader range of shapes and generalize our observations to dry systems, we complement the experiments with overdamped non-smooth contact dynamics simulations in which gear-shaped particles and crosses with characteristic size $d$ are driven at a constant velocity $V^{\mathrm{aff}}$ through the same channel geometry (Fig.~\ref{fig:fig1}(c-d); Supplementary Video~\textcolor{blue}{S3}). Consistent with experimental results, non-circular particles yield different scalings depending on shape (Fig.~\ref{fig:fig1}(e)), with $\ln (\savg + 1 )$ scaling linearly with $(w/d)^2$ for three-lobed gears, and sublinearly for crosses and six-lobed gears. 

\begin{figure*}
\includegraphics[width=1\linewidth]{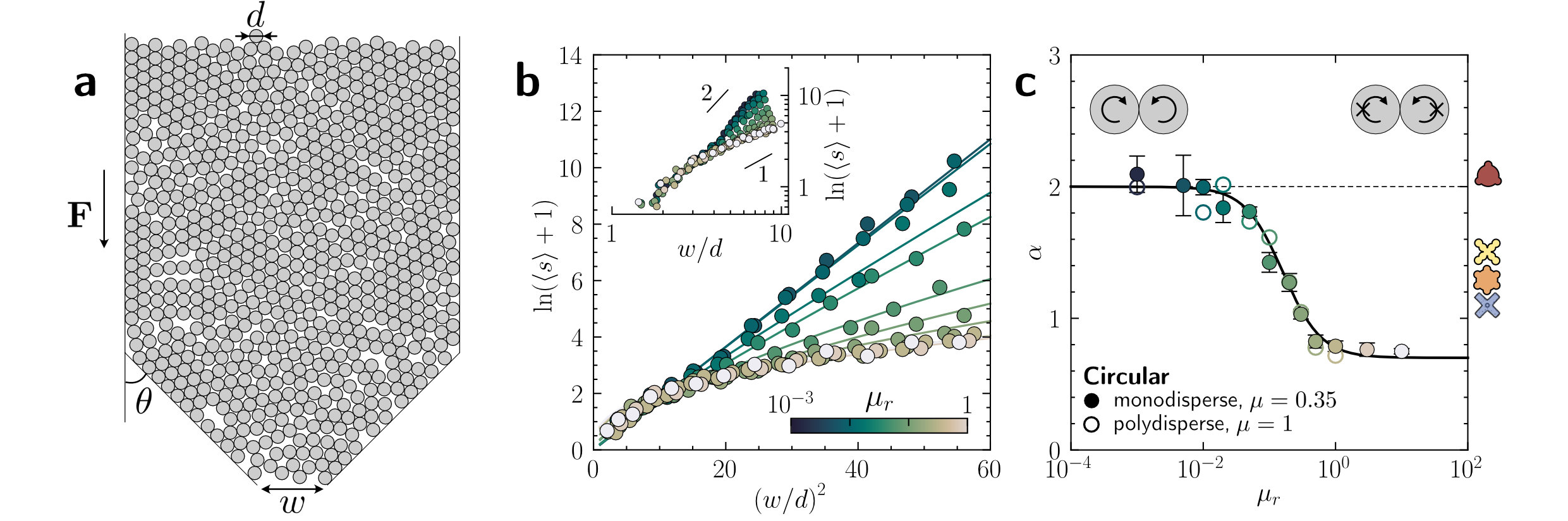}
\caption{\label{fig:fig2} \textbf{Influence of rolling friction on the average avalanche size.} \textbf{(a)} Molecular dynamics simulations of particles of diameter $d$ with sliding friction $\mu$ and rolling friction $\mu_r$ flowing through a hopper channel with constriction width $w$ and angle $\theta=45^{\circ}$ under a constant driving force $F$. \textbf{(b)} Avalanche size $\ln(\savg + 1)$ versus the squared size ratio $(w/d)^2$ for packings with varying $\mu_r$, fitted using equation~(\ref{eq:scaling}). The color scale indicates the rolling friction coefficient $\mu_r$. Inset: same data on a log-log scale. Solid black lines indicate slopes of 1 and 2 as guides to the eye. \textbf{(c)} Fitted power-law exponent $\alpha$ as a function of $\mu_r$ for monodisperse packings with $\mu=0.35$ (solid circles) and polydisperse packings with $\mu=1$ (open circles). Error bars represent the jackknife standard deviation of the power-law exponent. Symbols on the right indicate the values of $\alpha$ obtained for non-circular particles in experiments and simulations. The horizontal dashed line indicates predictions from equation~(\ref{eq:scaling}) in 2D, while the solid curve corresponds to a logistic fit.}
\end{figure*}

Together with the existing literature, our results confirm a clear shape-dependent deviation from equation~(\ref{eq:scaling}) in both experiments and simulations. To understand the physical origin of this deviation, we qualitatively compare the kinematics of circular and non-circular particles during hopper discharge. While circular particles easily rearrange at the channel outlet, crosses and gears form interlocked contacts that strongly constrain their relative motion. Such reduced mobility has also been reported during the discharge of 3D crosses and cubes \cite{hafez_effect_2021}, as well as in sheared non-circular packings \cite{saint-cyr_rheology_2011,azema_rheology_2013}. It is commonly captured numerically by simulating circular grains with resistance to both sliding and rolling \cite{ai_assessment_2011,estrada_identification_2011,singh_shear_2020}.

To isolate the role of pairwise constraints in the formation of clogs, we perform molecular dynamics (MD) simulations of 2D hopper flow, where circular particles with sliding and rolling friction are driven through a constriction by a constant force $F$ (Fig.~\ref{fig:fig2}(a)). Simulations are conducted on both monodisperse and polydisperse packings in a damped regime to limit inertial effects at the channel outlet. Tangential forces are computed using a linear spring-dashpot model and a Coulomb criterion for sliding with a coefficient $\mu$. Rolling friction is implemented via a rolling torque governed by a Coulomb-like threshold with a coefficient $\mu_r$ \cite{luding_cohesive_2008}. Both $\mu$ and $\mu_r$ are identical for interparticle and particle-wall interactions. All simulations are first performed with a high sliding friction ($\mu\in\{0.35,1\}$) and a systematically varied rolling friction ($0 \leq\mu_r \leq 10$) to investigate the influence of rolling constraints.

Fig.~\ref{fig:fig2}(b) shows the evolution of $\ln(\savg+1)$ with $(w/d)^2$ for monodisperse systems with varying $\mu_r$. While low-$\mu_r$ curves follow the expected squared-exponential scaling, the data at large rolling friction exhibit a strong deviation similar to that observed for non-circular particles. The power-law exponent $\alpha$ in equation~(\ref{eq:scaling}) and its associated error are estimated for each curve by extracting the log-log slope using a jackknife resampling method (inset of Fig.~\ref{fig:fig2}(b)). Although precise values of $\alpha$ at high $\mu_r$ may be biased by the limited size of avalanches, reflecting the transient discharge of grains at flow onset rather than the steady-state clogging probability (Supplementary Fig.~\textcolor{blue}{S1}), the data reveal a clear crossover from $\alpha=2$ at low $\mu_r$ to $\alpha \lesssim 1$ at high $\mu_r$ (Fig.~\ref{fig:fig2}(c)). This crossover is robustly observed in both monodisperse and polydisperse systems with large sliding friction.

The exponents $\alpha$ are also extracted for non-circular particles in experiments and simulations, yielding values ranging from $\alpha = 2.1 \pm 0.3$ for three-lobed gears to $\alpha = 1.1 \pm 0.1$ for experimental crosses. Experimental disks are omitted due to the limited number of data points at large $w/d$. These results confirm similar shape- and friction-dependent scalings of the clogging probability, with $\alpha \approx 2$ for roughly circular and low-$\mu_r$ particles, and $\alpha < 2$ for various non-circular and high-$\mu_r$ particles. The correspondence between shape and friction further suggests that both influence clogging through their common effect on relative particle mobility.

\begin{figure}[t]
\includegraphics[width=1\linewidth]{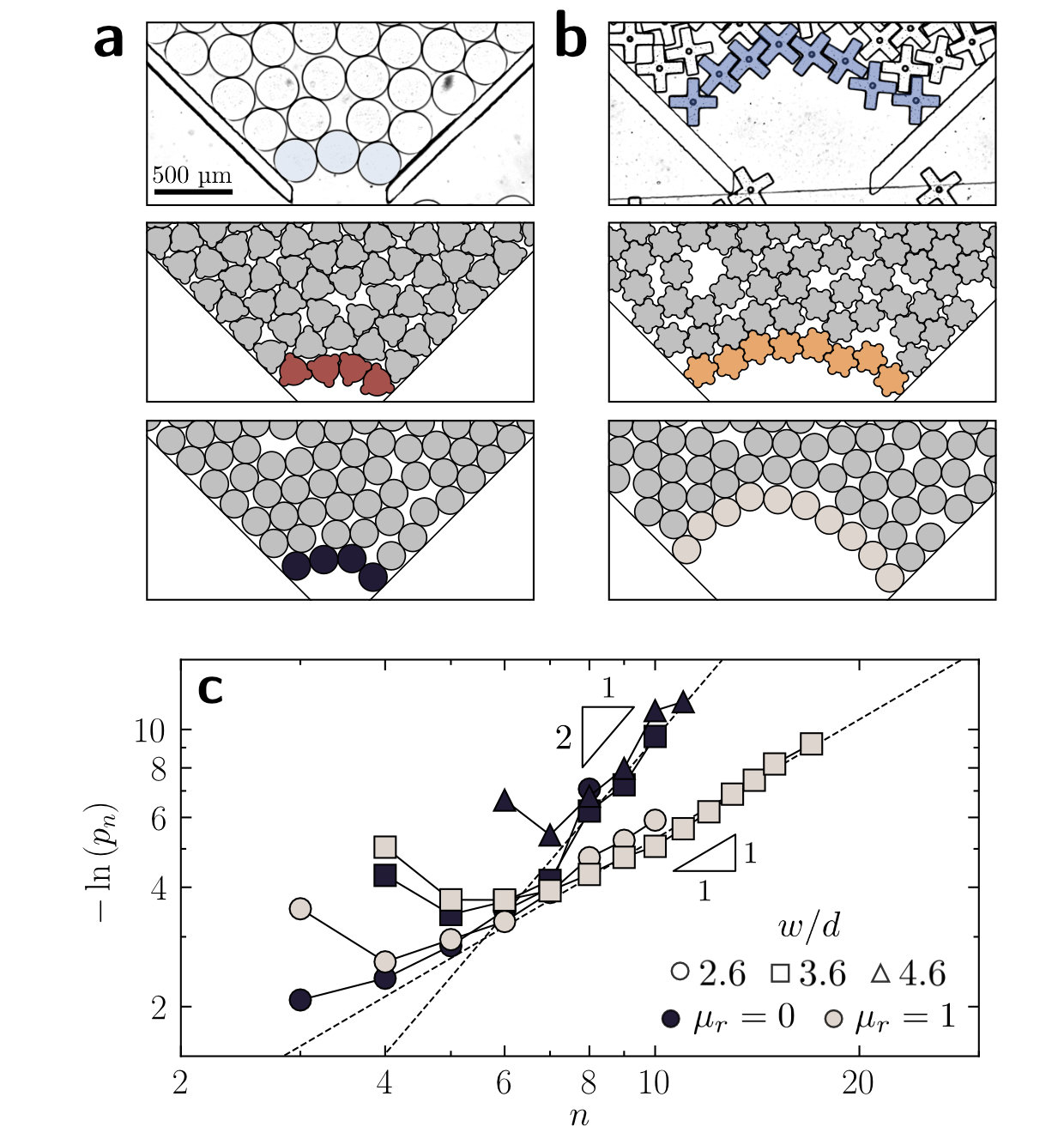}
\caption{\label{fig:fig3} \textbf{Influence of shape and rolling friction on the distribution of clogging arch sizes.} \textbf{(a)} Snapshots of clogging arches comprising relatively few particles, showing experimental disks, three-lobed gears, and low-$\mu_r$ particles (top to bottom). \textbf{(b)} Snapshots of large clogging arches, showing experimental crosses, six-lobed gears, and high-$\mu_r$ particles (top to bottom). \textbf{(c)} Arch size distributions $-\ln(p_n)$ versus $n$ at various size ratios $w/d$ for frictional circular particles ($\mu=0.35$) with $\mu_r=0$ (dark blue circles) and $\mu_r=1$ (beige circles).}
\end{figure}

This distinction between two regimes is also observed when comparing the distribution of clogging arch sizes. While circular, three-lobed, and low-$\mu_r$ grains generally form arches composed of relatively few particles (Fig.~\ref{fig:fig3}(a)), crosses, six-lobed, and high-$\mu_r$ grains exhibit a marked over-representation of large clogging arches (Fig.~\ref{fig:fig3}(b)). The difference in arch size is quantified by estimating the probability $p_n$ that an exiting particle triggers an arch of size $n$ at a fixed $w/d$. This probability is calculated as $p_n = p f(n)$, where $p= 1 / (\savg + 1)$ is the overall clogging probability and $f(n)$ is the fraction of clogs comprising $n$ particles.

Fig.~\ref{fig:fig3}(c) shows the distributions of $-\ln (p_n)$ at various size-ratios $w/d$ for circular particles with free ($\mu_r=0$) and blocked ($\mu_r=1$) rolling, each computed from at least 1,000 independent clogs. Depending on the rolling friction coefficient, the data for varying $w/d$ collapse onto two distinct scalings, with $-\ln(p_n) \sim n^2$ for $\mu_r=0$, and $-\ln(p_n)\sim n$ for $\mu_r=1$. Analogous shape-dependent scalings are observed for three-lobed and six-lobed gears (Supplementary Fig.~\textcolor{blue}{S2}). Deviations from these scalings for $n$ close to $w/d$ are explained by the cutoff in arch span imposed by the constriction width, which geometrically excludes certain clogging configurations \cite{to_jamming_2001}.
These results demonstrate that the scaling of the clogging probability with arch size exhibits a dependence on rolling resistance similar to that observed when varying the size ratio $w/d$. The collapse of $p_n$ for different $w/d$ further suggests that the probability of clog formation is primarily controlled by the arch size $n$ rather than $w/d$ itself, with the dependence of $\savg$ on the size ratio arising from the linear increase in average arch size with $w/d$ (Supplementary Fig.~\textcolor{blue}{S3}).

\begin{figure*}[t]
\includegraphics[width=1\linewidth]{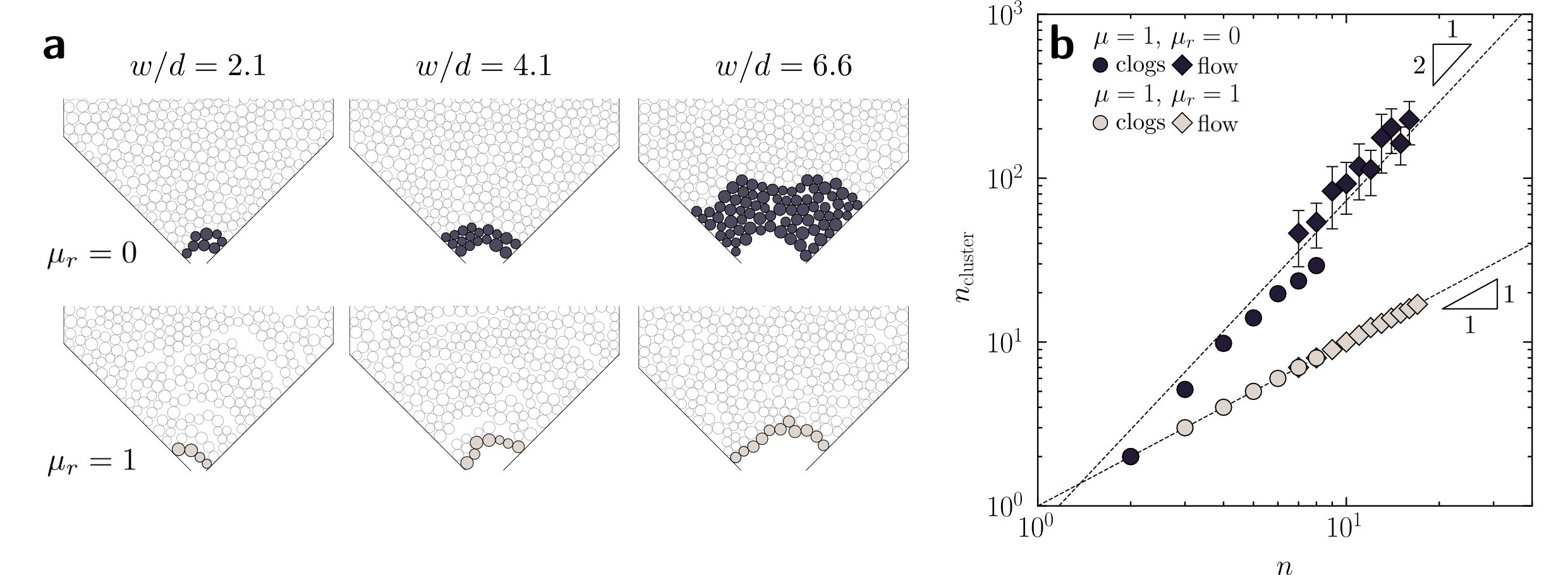}
\caption{\label{fig:fig4}\textbf{Minimal clogging cluster size for particles with free and blocked rolling.} \textbf{(a)} Representative examples of minimal clogging clusters in polydisperse packings with blocked sliding ($\mu=1$) and either free ($\mu_r=0$, top) or blocked ($\mu_r=1$, bottom) rolling. Colored particles constitute the rigid cluster, while white particles have been removed from the system. \textbf{(b)} Average minimal clogging cluster size $\nclus$ versus clogging arch size $n$ for $\mu_r=0$ (dark blue symbols) and $\mu_r=1$ (beige symbols) in clogs (circles) and during flow (diamonds). Dashed lines show a quadratic fit $\nclus \propto n^2$ and the linear relation $\nclus = n$. Error bars indicate 95\% confidence intervals of the average cluster size within each bin.}
\end{figure*}

To understand the dependence of the clogging probability on rolling friction, we must establish the physical conditions necessary for an arch of $n$ particles to form a clog. For clogging to occur, a spanning group of particles must achieve mechanical equilibrium during discharge. However, equilibrium alone is insufficient; the structure must also be globally stable to prevent arch collapse under slight perturbations. The stability of disordered systems is generally analyzed using rigidity theory. An assembly of repulsive particles at mechanical equilibrium is stable if it possesses no internal zero-energy displacement modes, or floppy modes. Such a state is called rigid. The number $N_0$ of these floppy modes is determined by a balance between the number of internal degrees of freedom $N_{\mathrm{free}}$ and the number of constraints $N_{\mathrm{cons}}$ in the system, as formalized by the Maxwell-Calladine count \cite{maxwell_ireciprocal_1870,calladine_buckminster_1978}:
\begin{equation}
    N_0 + f(D) = N_{\mathrm{free}} - N_{\mathrm{cons}} + N_{s},
    \label{eq:maxwell}
\end{equation}
where $f(D)=D(D+1)/2$ is the number of rigid-body translations and rotations under free boundary conditions in $D$ dimensions, and $N_{s}\geq0$ is the number of states of self-stress, which correspond to redundant constraints not needed for rigidity.

We apply this framework to assess the rigidity of a spanning arch of $n$ particles. In 2D, each particle possesses three degrees of freedom (two translational and one rotational), yielding $N_{\mathrm{free}}=3n$. In the absence of rolling resistance, and assuming no contact can slide, each of the $n-1$ interparticle contacts and $2$ particle-wall contacts imposes two constraints (normal and tangential), leading to $N_{\mathrm{cons}}=2(n+1)$. Moreover, since the arch is pinned to fixed boundaries, no rigid-body motion of the system is possible, such that $f(D)=0$. Equation~(\ref{eq:maxwell}) therefore gives $N_0 = n-2 + N_{s} \geq n-2$, demonstrating that any arch of freely rolling frictional particles larger than $n=2$ has floppy modes and is mechanically unstable. 

To achieve rigidity, such an arch requires additional particles to constrain its $\mathcal{O}(n)$ floppy modes. Assuming the flowing bulk sits at or below generalized isostaticity, where each particle adds on average enough contacts to balance its own degrees of freedom~\cite{shundyak_force_2007}, extra constraints must be supplied by contacts with the channel walls. Eliminating these floppy modes therefore requires $\mathcal{O}(n)$ wall contacts, corresponding to roughly $n$ layers of particles above the arch. Consequently, the total number of particles involved in clog stability is expected to scale as $n^2$. Considering granular discharge as a stochastic sampling of flow configurations, where each of these $\sim n^2$ particles has a probability $q$ of being in a specific position-force microstate~\cite{thomas_fraction_2015}, the probability of sampling a clogging configuration scales as $p_n \sim \exp \left[-n^2 \ln (1/q) \right]$, recovering the observed scaling for $\mu_r=0$.

Introducing rolling resistance fundamentally alters this constraint counting. While the number of degrees of freedom remains $N_{\mathrm{free}}=3n$, each contact below the Coulomb threshold for rolling now imposes an additional rotational constraint, leading to $N_{\mathrm{cons}}=3(n+1)$ and $N_0 = N_s - 3$. Because a simple chain of particles contains no internal topological redundancy, redundant constraints arise solely from the arch contacting two walls whereas a single wall contact would suffice to achieve rigidity. This gives $N_s = 3$ and $N_0 = 0$, demonstrating that any spanning arch of $n$ particles with blocked sliding and rolling is inherently rigid. Because no additional particles are required to achieve stability, the total number of particles involved in clog formation is expected to scale as $n$, yielding $p_n \sim \exp \left[-n \ln (1/q) \right]$, as observed for $\mu_r=1$.

Verifying these scaling arguments requires directly measuring the size of the clogging region and its dependence on rolling friction. To this end, we determine the minimum number of particles required for arch rigidity in MD simulations using the pebble game algorithm 
\cite{jacobs_generic_1995,jacobs_algorithm_1997}, which decomposes the contact network into rigid and floppy regions based on a balance between local constraints and internal degrees of freedom. Specifically, we apply the (3,3) variant of the algorithm, which enables the detection of rigid clusters in the presence of sliding \cite{henkes_rigid_2016,liu_frictional_2019,van_der_naald_minimally_2024} and rolling \cite{lester_frictional_2018} friction (Supplementary Video~\textcolor{blue}{S4}). 
For a given arch within a rigid cluster, the minimum number of particles required for clogging is determined by iteratively removing particles from the system until all remaining particles are strictly necessary to preserve arch rigidity (see Methods and Supplementary Video~\textcolor{blue}{S5}). We define the resulting subset of particles as the minimal clogging cluster. To quantify the influence of interparticle rolling on the clogging cluster size, we apply the pebble game to polydisperse packings with blocked sliding ($\mu=1$) and either free ($\mu_r=0$) or blocked ($\mu_r=1$) rolling. Rigid clusters are identified both in clogged configurations and at regular time intervals during flow (Supplementary Fig.~\textcolor{blue}{S4}).

Examples of minimal clogging clusters obtained for various outlet-to-particle size ratios in the absence and presence of rolling friction are shown in Fig.~\ref{fig:fig4}(a). While minimal clusters reduce to the spanning arch when both sliding and rolling are blocked (bottom row), freely rolling particles require a stabilizing region whose area appears to increase with the arch size and constriction width (top row). The average cluster size $\nclus$ is plotted against the arch size $n$ in Fig.~\ref{fig:fig4}(b). Consistent with our scaling arguments, the minimal cluster size simply equals $n$ when both sliding and rolling are blocked, as no additional constraints are needed for rigidity. In contrast, allowing particles to roll freely leads to the predicted quadratic relationship, $\nclus \propto n^2$, both on clogs and during flow. The slight offset between clogged and flowing states arises from the lower average contact number per particle during discharge (Supplementary Fig.~\textcolor{blue}{S4}). The same result is obtained independently of the hopper angle (Supplementary Fig.~\textcolor{blue}{S5}). These two scalings closely match the observed variations in avalanche size with the size ratio $w/d$ and the arch size $n$. This strongly suggests that the clogging probability is fundamentally controlled by the number of particles required for arch rigidity, which is determined by a balance between constraints and degrees of freedom. 

The dependence of clogging on local constraints implies that any property restricting interparticle mobility should affect the scaling of the clogging probability. This applies not only to factors that hinder rotation, but also to those that limit relative sliding. Indeed, from a rigidity standpoint, rotational and tangential constraints are equivalent, as each restricts a single degree of freedom. A system with blocked sliding and a given fraction of rolling contacts is therefore expected to behave similarly to a system with blocked rolling and an equal fraction of sliding contacts. This equivalence can be directly derived in MD simulations from the formulation of the sliding and rolling criteria. Two contacting particles of radius $R$ slide when the tangential force reaches $\mu F_n$, where $F_n$ is the normal force, inducing a torque $\tau_t = R \mu F_n$. Similarly, rolling occurs when the rolling torque $\tau_r$ reaches $\tau_r = (R/2) \mu_r F_n$, where $R/2$ is the effective particle radius (see Methods). We therefore expect systems with blocked rolling and a sliding friction $\mu$ to exhibit the same power-law exponent $\alpha$ as systems with blocked sliding and a rolling friction $\mu_r=2\mu$.

\begin{figure}[t]
\includegraphics[width=1\linewidth]{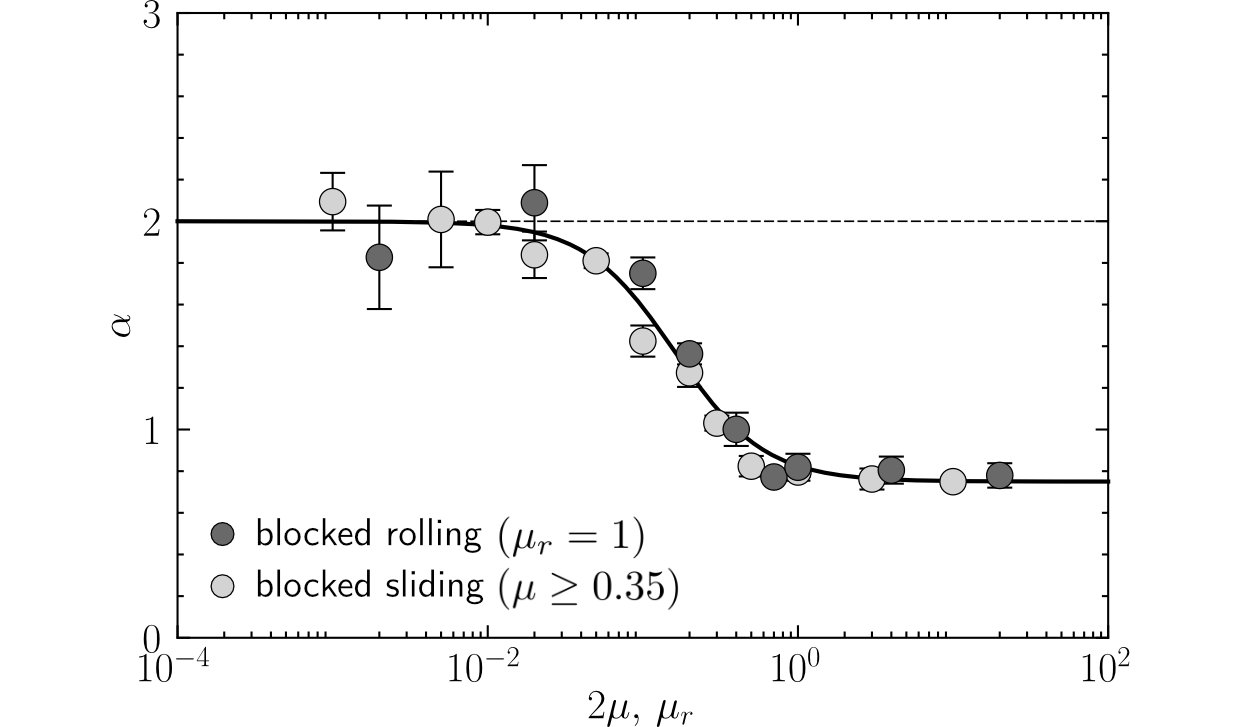}
\caption{\label{fig:fig5}\textbf{Influence of sliding and rolling friction coefficients on $\alpha$.} Fitted power-law exponent $\alpha$ as a function of $\mu_r$ for particles with blocked sliding ($\mu\geq 0.35$, white circles), and as a function of $2\mu$ for particles with blocked rolling ($\mu_r = 1$, gray circles). Error bars represent the jackknife standard deviation of $\alpha$. The horizontal dashed line indicates predictions from equation~(\ref{eq:scaling}), while the solid curve corresponds to the same logistic fit as in Fig.~\ref{fig:fig2}(c).}
\end{figure}

We verify this equivalence by performing additional MD simulations on monodisperse packings with $\mu_r = 1$ and varying sliding friction coefficients $\mu$. The fitted exponents $\alpha$ are plotted against $2\mu$ for high rolling friction ($\mu_r = 1$), and against $\mu_r$ for high sliding friction ($\mu \geq 0.35$) in Fig.~\ref{fig:fig5}. As predicted, all data collapse onto a single curve, demonstrating that the ability of particles to both slide and roll dictates the scaling of the clogging probability with the ratio $w/d$. These results further confirm that clog formation is primarily controlled by the relative mobility between particles, which ultimately depends on friction and shape.

In this paper, we establish a direct connection between the microscopic properties of grains and the macroscopic probability of clogging in confined geometries. Combining microfluidic experiments, numerical simulations, and rigid cluster analysis, we demonstrate that particle shape and friction determine the minimum number of particles required to form a rigid spanning structure through the balance between degrees of freedom and interparticle constraints. Overall, our results provide a physical foundation for the long-standing empirical scaling of the clogging probability with the outlet-to-particle size ratio and explain its breakdown for non-circular particles.

Although this study focuses on 2D systems, our framework naturally extends to 3D assemblies of freely sliding or rolling spheres. By analogy with 2D arches, a 3D clogging dome comprising $n^2$ particles possesses $\mathcal{O}(n^2)$ floppy modes. Because each particle layer contacts the channel wall along a perimeter proportional to $n$, eliminating all floppy modes requires roughly $n$ layers of particles above the dome. The total number of particles involved in clog stability therefore scales as $n^3$, consistent with the scaling laws reported in 3D \cite{thomas_fraction_2015,montero_dimensional_2025}. Directly measuring the minimum clogging cluster size in 3D remains a challenge, as current combinatorial algorithms such as the pebble game are restricted to 2D systems \cite{chubynsky_algorithms_2007}.

More broadly, our results imply that predicting flow arrest in confined geometries requires considering not only the conditions for mechanical equilibrium, but also those for rigidity. As a consequence, any property that modifies the balance between degrees of freedom and constraints is expected to significantly alter the clogging probability. This applies to properties that add degrees of freedom, such as deformability, or modify interparticle constraints, such as friction and shape. It also extends to properties that affect the connectivity of the contact network, such as dilution and inertia. Altogether, our findings provide new physical insights for understanding and controlling clog formation in confined granular flows.

\putbib[bibli]
\end{bibunit}

\clearpage

\begin{bibunit}

\section*{Methods}
\subsection*{Microfluidic experiments}
Each discharge experiment is performed in a PDMS (polydimethylsiloxane) channel consisting of a $24$ mm long straight section of width $5$ mm and thickness $92$ \textmu m, followed by a tapered section at $45^{\circ}$ and a constriction of width $w=600$ or $900$ \textmu m \cite{kool_hopper_2025}. A pneumatically controlled Quake valve \cite{unger_monolithic_2000} is placed immediately after the constriction, retaining particles while allowing for fluid flow when closed. Particles are fabricated using an \textit{in situ} photo-lithographic projection method \cite{dendukuri_stop-flow_2007, cappello_transport_2019}. The channel is first filled with a UV-polymerizable solution consisting of 99.5\% v/v Poly(ethylene glycol) diacrylate (PEGDA, $M_N$ = 575), mixed with 0.5\% v/v photo-initiator (Darocur 1173). Upon exposure to UV light ($\lambda = 365\pm 10$ nm) during a time $t=1$ s, the solution locally cross-links and forms a solid hydrogel-like particle. The shape and size of the particles are controlled by placing a black plastic mask with a transparent shape at its center in the field-stop position of the microscope, using a 10x objective. We focus on circular particles with diameter $270 \leq d \leq 379$~\textmu m, and crosses with arm length $a\in \{100,200\}$~\textmu m (arm width $a/2$). The microfluidic channel is filled with hundreds of particles by repeating the aforementioned UV cross-linking method and moving the focal spot of the objective between successive UV pulses. Fabricated particles are then packed at the constriction of the hopper by closing the Quake valve while imposing a fluid flow.

Once the particles are densely packed at the constriction, the Quake valve is opened while maintaining flow of the un-crosslinked solution, allowing the suspension to discharge through the channel constriction. Particles are driven by a constant pressure drop $\Delta P = 10$ mbar between the inlet and outlet of the channel. If a clog forms and remains stable for more than $4$ seconds, a brief reverse flow is applied to break the obstructing arch and restore particle flow. Experiments end when the number of remaining particles is insufficient to fill the tapered section of the channel. Each experiment is recorded using a digital CMOS camera capturing at $5$ Hz. Particle positions are extracted using a custom spatial convolution algorithm, and individual trajectories are reconstructed using the open-source tracking algorithm Trackpy \cite{allan_soft-mattertrackpy_2024}. To ensure the statistical convergence of the average avalanche size, approximately $30$ clogs are recorded for each particle shape and constriction width.

\subsection*{Non-smooth contact dynamics simulations}

Simulations of driven overdamped frictionless hard particles are performed using the event-driven scheme introduced in~\cite{lerner_simulations_2013}. For a system of frictionless hard disks, the geometry of the instantaneous contact network is encoded in the operator
\begin{equation}
S \equiv \frac{\partial r_{ij}}{\partial \mathbf{x}_k},
\end{equation}
where \(r_{ij}\) denotes the distance between particles $i$ and $j$, and $\mathbf{x}_k$ is the position of particle $k$. Each particle $i$ has a diameter $d_i$. The hard-core constraint, which prevents overlaps between particles already in contact, imposes
\begin{equation}
S |V\rangle = 0,
\end{equation}
where $|V\rangle$ is the vector of particle velocities. In the overdamped limit, force balance reads
\begin{equation}
|F^{\mathrm{drag}}\rangle
+
|F^{\mathrm{cont}}\rangle
=0,
\end{equation}
with drag and contact forces given by
\begin{equation}
|F^{\mathrm{drag}}\rangle = -\xi_0^{-1}(|V\rangle-|V^{\rm aff}\rangle),
\qquad
|F^{\mathrm{cont}}\rangle = S^{T}|f\rangle,
\end{equation}
where $|f\rangle$ is the vector of contact forces, $\xi_0^{-1}$ is the drag coefficient, and $|V^{\rm aff}\rangle$ are the affine velocities that drive the particles in the channel. The particle velocities therefore satisfy
\begin{equation}
|V\rangle
=
|V^{\rm aff}\rangle
+
\xi_0 S^{T}|f\rangle .
\end{equation}
Applying the contact constraint $S|V\rangle = 0$ yields
\begin{equation}
|f\rangle
=
-\xi_0^{-1}
(SS^{T})^{-1}
S |V^{\rm aff}\rangle .
\end{equation}
Defining the contact matrix $\mathcal{N} \equiv SS^{T}$, the particle velocities can be written as
\begin{equation}
|V\rangle
=
\left(
I - S^{T}\mathcal{N}^{-1}S
\right)
|V^{\rm aff}\rangle .
\end{equation}
In the present work, particles are driven along the horizontal direction through the affine flow
\begin{equation}
|V^{\rm aff}\rangle = Q |e_x\rangle,
\end{equation}
where \(Q\) sets the driving amplitude and $|e_x\rangle$ is the unit vector along the channel axis. The dynamics evolve between collision events while enforcing the constraints
\begin{equation}
f_{ij} > 0,
\qquad
r_{ij} \ge \frac{d_i + d_j}{2} .
\end{equation}
Whenever a collision generates negative contact forces, the contact network is rewired by iteratively removing contacts with \(f_{ij}<0\) until all constraints are satisfied. Further details of the algorithm can be found in~\cite{lerner_simulations_2013}.

We extend this framework by including permanent contacts to construct particles with non-spherical rigid shapes. Additional redundant constraints are introduced such that each composite particle behaves as a rigid body, generating a state of self-stress in $\mathcal{N}$. The same hopper geometry as in experiments is implemented by applying a harmonic restoring force on the side walls of the straight section, and adding two rigid segments to represent the walls in the tapered section. Consequently, the matrix $\mathcal{N}$ has dimensions $m \times m$, where $m$ is the total number of constraints in the system, namely: the number of permanent contacts defining each rigid particle, the number of non-permanent disk--disk contacts between disks belonging to different composite particles, and the number of active disk--segment contacts at a given time.

This work focuses on two particle geometries: gear-like particles and four-fold crosses. Gear-like particles are composed of one large disk of diameter $d_l$ surrounded by $n_{\mathrm{lobes}}$ smaller disks of diameter $d_s=0.35d_l$, uniformly distributed along the perimeter of the central disk with angular spacing $2\pi/n_{\mathrm{lobes}}$. The resulting effective gear diameter is $d = d_l + d_s$. Four-fold crosses are constructed from nine disks: one central disk and four arms, each arm consisting of two aligned disks. The number of particles is set to $N=700$ and $1000$ for gears and crosses, respectively. All particles exiting the channel are reinjected at random positions at the back of the hopper.

\subsection*{Molecular dynamics simulations}

Molecular dynamics (MD) simulations are performed using LAMMPS \cite{thompson_lammps_2022}. A 2D hopper channel of width $L=50$ and constriction angle $\theta=45^{\circ}$ is filled with $N=1000$ circular particles of mean diameter $\bar d=2$ and density $\rho=2.4\times 10^{-3}$ (dimensionless units). The particle diameters are either strictly identical or uniformly distributed between $0.8\bar d$ and $1.2\bar d$ to form a packing with a polydispersity of $\sigma \approx 11.6\%$. All particles are driven towards the constriction by a constant external force $F=1$ applied along the channel axis. To remain in a damped flow regime, particles are also subjected to a viscous drag force $\mathbf{F}_v = - \eta \mathbf{v}$, where $\eta=1$ is the drag coefficient and $\mathbf{v}$ is the particle velocity. Particles escaping the channel are reinserted at random positions upstream of the packing, where a buffer distance is enforced to prevent artificial contacts across the periodic domain. A clog is detected when the average kinetic energy of the system falls below a threshold of $10^{-9}$. All particles are then removed and a new simulation is started, ensuring consistent initial conditions for every run.

Particle-particle and particle-wall interactions are described using a soft-sphere contact model that accounts for normal repulsion, tangential friction, and rolling resistance. Normal interactions are modeled using a linear spring-dashpot formulation with normal spring stiffness $k_n$ and damping coefficient $\gamma_n$. Friction forces are modeled using a linear spring-dashpot-slider model with a tangential stiffness $k_t$ and damping coefficient $\gamma_t$, bounded by a Coulomb sliding criterion with a coefficient $\mu$. Resistance to rolling is implemented by computing a rolling pseudo-force $F_r$ using a linear spring-dashpot-slider model with a rolling stiffness $k_r$ and damping coefficient $\gamma_r$, bounded by a Coulomb rolling criterion with a coefficient $\mu_r$ \cite{luding_cohesive_2008,ai_assessment_2011}. Although this pseudo-force does not contribute to the force balance, it induces a rolling torque of magnitude $\tau_r = R_{\mathrm{eff}}F_r$, where $R_{\mathrm{eff}}=R_iR_j/(R_i + R_j)$ is the effective radius for two contacting particles $i$ and $j$ \cite{luding_cohesive_2008}.
The sliding friction coefficient is fixed at $\mu=0.35$ unless otherwise specified, while the rolling friction coefficient is systematically varied from $\mu_r=0$ to $\mu_r=10$. The normal spring stiffness is set to $k_n=10^5$, sufficiently large to ensure that overlaps remain small compared to the particle size. The normal damping coefficient is set to $\gamma_n=2.1$, corresponding to a restitution coefficient of 0.86 \cite{shafer_force_1996}. Following~\cite{shafer_force_1996,luding_cohesive_2008}, the tangential and rolling stiffnesses are set to $k_t=2k_n/7$ and $k_r=0.1k_n$, while the tangential and rolling damping coefficients are set to $\gamma_t=0.2\gamma_n$ and $\gamma_r = 0.05\gamma_n$. All friction, stiffness, and damping coefficients are identical for both particle-particle and particle-wall interactions. 
The force network is extracted at regular timesteps, with each contact classified based on whether the Coulomb thresholds for sliding and rolling are reached.

\subsection*{Pebble game algorithm}

Rigid particle clusters are detected in molecular dynamics simulations using the pebble game algorithm, which decomposes the system into rigid and floppy regions based on a balance between internal degrees of freedom and independent constraints \cite{jacobs_generic_1995,jacobs_algorithm_1997}. The contact network is first mapped onto a 2D graph, where vertices correspond to particles, and edges represent the constraints imposed by each contact: one edge when particles are free to slide and roll, two when either sliding or rolling is blocked, and three when both are blocked. All walls of the system are treated as a single ground vertex, as the distance and angle between them are fixed by the channel geometry \cite{jacobs_algorithm_1997}. We then apply the $(3,3)$ variant of the algorithm \cite{henkes_rigid_2016} using the open-source repository by Henkes \cite{henkes_silkehenkesrigidlibrary_2026}, which we modified to incorporate triple constraints. In this variant, each vertex of the 2D graph is initially assigned three pebbles, representing its three degrees of freedom (two translational and one rotational). Each constraint is then classified as independent or redundant depending on whether the corresponding edge can be covered by four pebbles \cite{jacobs_algorithm_1997}. Regions with enough independent constraints to block all internal degrees of freedom are identified as rigid.

We apply the pebble game algorithm to identify the smallest subset of particles required to maintain the rigidity of a spanning arch at the constriction. For a given arch within a rigid cluster, this subset is obtained following an approach similar to that used to measure the cutting length of jammed systems \cite{wyart_geometric_2005,goodrich_stability_2013}. We iteratively remove randomly selected particles from the upper boundary of the rigid cluster until every remaining particle is strictly necessary to preserve arch rigidity. The resulting group of particles constitutes the smallest cluster capable of forming a clog. Although this method may yield slightly different clusters depending on the order of removal, we verify that the relative standard deviation of the cluster size remains below 5\% (Supplementary Fig.~\textcolor{blue}{S4}(a)).

Because the pebble game algorithm is only valid for generic networks with no spatial symmetries, minimal clusters are measured on packings with size polydispersity $\sigma \approx 11.6\%$. To avoid sampling bias when relating arch and cluster sizes, we identify rigid clusters in both flowing and clogged states. Analyzing only clogs would bias our measurements toward smaller clusters, which are more likely to form stable clogging structures. Conversely, analyzing only flowing states would exclude the smallest clusters that rapidly arrest flow, leading to an overestimation of the average cluster size (Supplementary Fig.~\textcolor{blue}{S4}(b)). Furthermore, flowing systems require on average more particles to form rigid spanning clusters due to the lower mean contact number per particle during discharge compared with static clogged configurations (Supplementary Fig.~\textcolor{blue}{S4}(c)). If no spanning rigid cluster is detected, which can occur in flowing states, no further analysis is performed. We focus on the two limiting regimes of blocked sliding with free rolling ($\mu=1$, $\mu_r=0$), and blocked sliding with blocked rolling ($\mu=1$, $\mu_r=1$). Intermediate regimes are harder to characterize, as the mobilization of contacts upon clogging does not reflect that of the flowing state.

\section*{Acknowledgments}
The authors thank Emanuela Del Gado for insightful discussions and for suggesting the use of the pebble game algorithm.

\section*{Funding}
This work was supported by the Institut Pierre-Gilles de Gennes (Investissements d'Avenir, grant ANR-10-EQPX-34). J.T. discloses support for the research of this work from ENS de Lyon through a CDSN grant.

\putbib[bibli]
\end{bibunit}

\clearpage
\onecolumngrid

\setcounter{equation}{0}
\setcounter{figure}{0}
\setcounter{table}{0}
\setcounter{section}{0}

\renewcommand{\theequation}{S\arabic{equation}}
\renewcommand{\thefigure}{S\arabic{figure}}
\renewcommand{\thetable}{S\arabic{table}}
\renewcommand{\thesection}{S\arabic{section}}

\renewcommand{\theHequation}{S\arabic{equation}}
\renewcommand{\theHfigure}{S\arabic{figure}}
\renewcommand{\theHtable}{S\arabic{table}}
\renewcommand{\theHsection}{S\arabic{section}}

\begin{bibunit}

\begin{center}
\textbf{\large Supplementary Information}
\end{center}
\vspace{0.5cm}

\section*{Supplementary Videos}

\noindent\textbf{Supplementary Video S1: Discharge of disks through a 2D microfluidic hopper.} \\
Microfluidic experiment in which disks of diameter $d=257$~\textmu m are discharged through a constriction of width $w=900$~\textmu m under imposed pressure.
\vspace{.5cm}

\noindent\textbf{Supplementary Video S2: Discharge of crosses through a 2D microfluidic hopper.} \\
Microfluidic experiment in which crosses with arm length $a=120$~\textmu m are discharged through a constriction of width $w=900$~\textmu m under imposed pressure.
\vspace{.5cm}

\noindent\textbf{Supplementary Video S3: Simulated hopper discharge of 6-lobed gears.} \\
Non-smooth contact dynamics simulation of 6-lobed gears driven at a constant velocity $V^{\mathrm{aff}}$ through a constriction of width $w=5.3d$, where $d$ is the effective gear diameter. The clogging arch is highlighted in dark gray once formed.
\vspace{.5cm}

\noindent\textbf{Supplementary Video S4: Rigid clusters during hopper discharge.} \\
Molecular dynamics simulation ($\mu=1$, $\mu_r=0$, $w/d=6.1$) with rigid clusters detected using the pebble game algorithm. Each colored group of contacting particles represents a rigid cluster. A clog forms when a spanning rigid cluster satisfies mechanical equilibrium.
\vspace{.5cm}

\noindent\textbf{Supplementary Video S5: Identification of the minimal clogging cluster.} \\
Starting with a configuration obtained from a molecular dynamics simulation ($\mu=1$, $\mu_r=0$, $w/d=6.1$), particles are iteratively removed from the upper boundary of the system while applying the pebble game algorithm until every remaining particle is strictly necessary to preserve arch rigidity. The clogging arch is highlighted in dark gray, while removed particles are shown in white. Colored networks indicate rigid clusters.
\vspace{.5cm}

\begin{figure}[h!]
\includegraphics[width=1\linewidth]{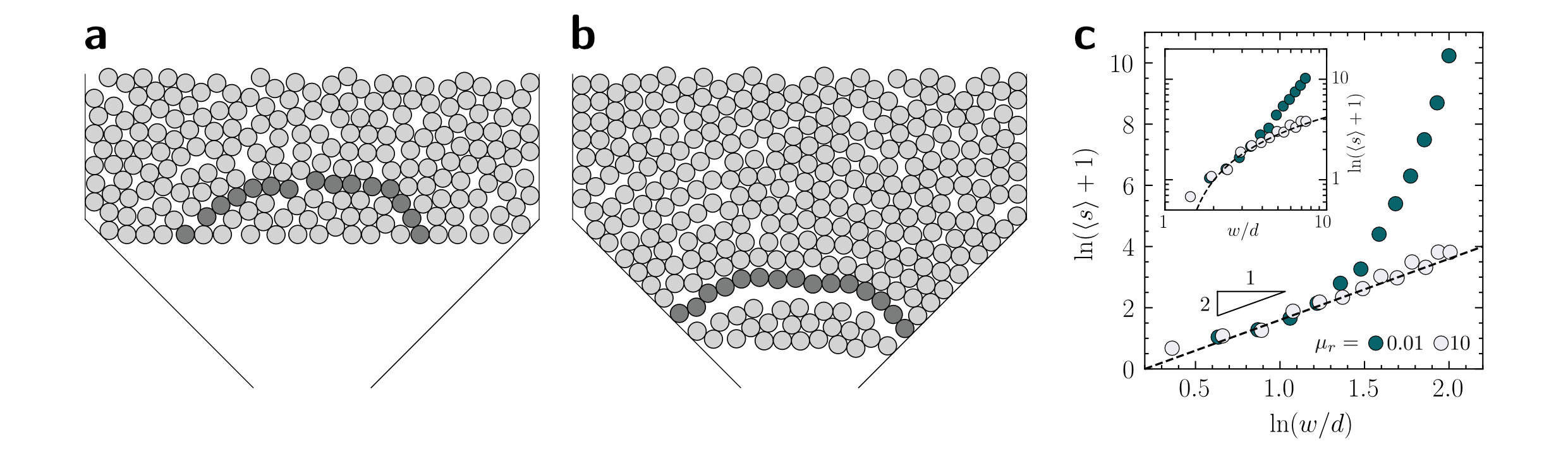}
\caption{\label{fig:SIfig_mins} \textbf{Transient avalanches at flow onset for large $\mu_r$ particles.} \textbf{(a-b)} Arch formation at flow onset for particles with large sliding ($\mu=0.35$) and rolling ($\mu_r=1$) friction coefficients. Dark gray particles form the clogging arch. Particles below the arch determine the avalanche size. \textbf{(c)} Dependence of $\ln(\savg + 1)$ on $\ln(w/d)$ for particles with $\mu=0.35$ and $\mu_r = 0.01$ (dark green circles) or $\mu_r = 10$ (white circles). While the low-$\mu_r$ data do not exhibit a linear trend, the high-$\mu_r$ data follow a linear scaling with a fitted slope of $2.0 \pm 0.1$ (dashed black line), characteristic of transient discharge at flow onset. Inset: The same data plotted as $\ln (\savg + 1)$ versus $w/d$ on logarithmic axes.}
\end{figure}

\section{Avalanche size at large rolling friction}

During granular discharge, the probability $p$ that a given grain triggers a clog is inferred from the average avalanche size $\savg$ via the relation $p = 1 / (\savg + 1)$, assuming a steady-state flow with a constant clogging probability. While this assumption holds for systems with relatively low rolling friction, where avalanches are sufficiently large to reach a steady state, it breaks down when both sliding and rolling friction are large. As described later in the main text, systems with blocked sliding and rolling can clog as soon as a spanning chain of particles forms. In such cases, the measured avalanche size reflects the transient discharge of grains at flow onset rather than the steady-state clogging probability. This is illustrated in Fig.~\ref{fig:SIfig_mins}(a-b), where an arch of particles forms immediately upon flow onset, such that the avalanche size corresponds solely to the number of particles already lying below the arch when the simulation begins.
Because clogging arches are generally convex \cite{to_jamming_2001,garcimartin_shape_2010}, the region occupied by these particles can be approximated by a semicircle of diameter $w$, following $\savg \sim (w/d)^2$. This scaling is confirmed in Fig.~\ref{fig:SIfig_mins}(c), where a linear fit of $\ln(\savg + 1)$ versus $\ln(w/d)$ yields a slope of $2.0 \pm 0.1$. Consequently, the dependence of $\ln(\savg+1)$ on $w/d$ measured numerically at large $\mu$ and $\mu_r$ likely does not obey the steady-state scaling of equation~(\textcolor{blue}{1}), rendering the fitted values of $\alpha$ unreliable in this range of parameters.

\section{Arch size distribution for non-circular particles}

\begin{figure}[h!]
\includegraphics[width=1\linewidth]{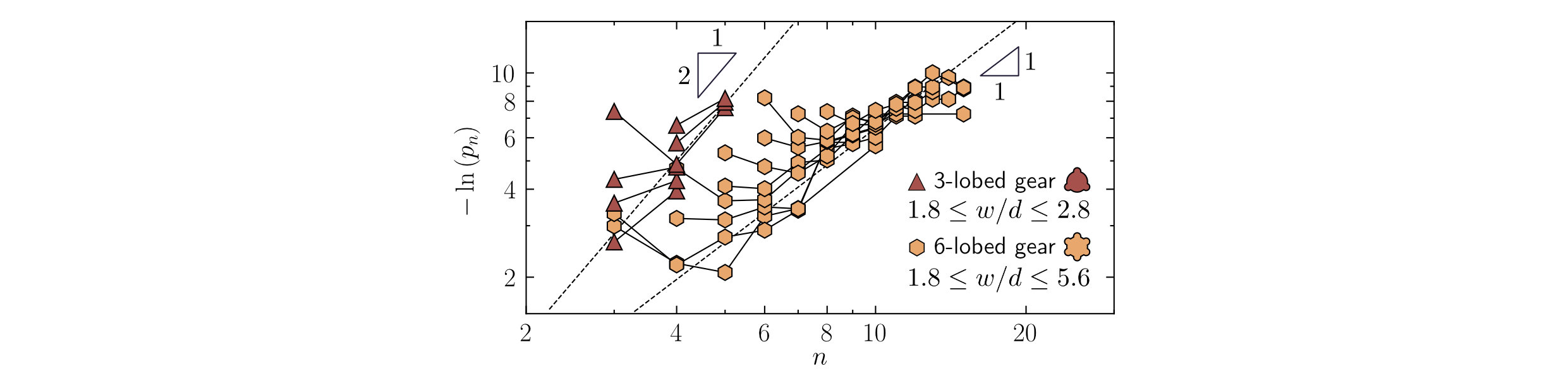}
\caption{\label{fig:SIfig_arch_size} \textbf{Distribution of clogging arch sizes for non-circular particles.} Arch size distributions $-\ln(p_n)$ versus $n$ at various size ratios $w/d$ for three-lobed (red triangles) and six-lobed (orange hexagons) gears. Dashed lines indicate linear and quadratic scalings.}
\end{figure}

While the main text focuses on arch size distributions of frictional circular particles with free ($\mu_r=0$) and blocked ($\mu_r=1$) rolling, analogous distributions can be obtained for non-circular particles in non-smooth contact dynamics simulations. Fig.~\ref{fig:SIfig_arch_size} shows the arch size distributions obtained for three-lobed and six-lobed gears. Although the statistical sample is smaller than that of circular particles in MD simulations, the data support the existence of shape-dependent scalings, with three-lobed gears exhibiting a distribution closer to $-\ln(p_n) \sim n^2$, whereas six-lobed gears exhibit a distribution closer to $-\ln(p_n)\sim n$. This difference is attributed to the weaker geometrical constraints of three-lobed gears, which restrict interparticle sliding and rolling less strongly. Experimental arch size distributions are omitted due to a limited amount of data.

\section{Relationship between arch size and constriction width}

\begin{figure}
\includegraphics[width=1\linewidth]{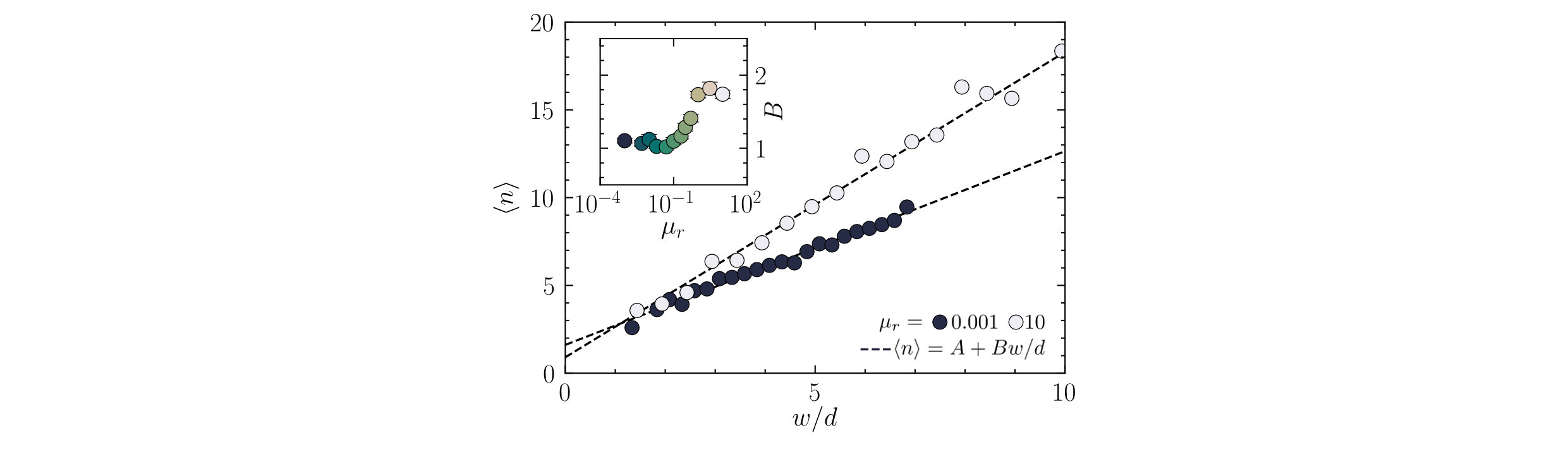}
\caption{\label{fig:SIfig_n_w/d} \textbf{Linear scaling between arch size and constriction width.} Dependence of the average arch size $\langle n \rangle$ on the outlet-to-particle size ratio $w/d$ for particles with sliding friction $\mu=0.35$ at low and high rolling friction $\mu_r$. The black dashed lines correspond to linear fits.  Inset: The slope of this linear dependence as a function of $\mu_r$, demonstrating a crossover from approximately $1$ at low $\mu_r$ to approximately $1.8$ at large $\mu_r$.}
\end{figure}

The clogging probability yields similar scaling exponents when plotted as a function of the outlet-to-particle size ratio $w/d$ and the arch size $n$. This correspondence is explained by the linear relationship between constriction width and the number of particles needed to span the channel, which maps the scaling in $n$ onto a scaling in $w/d$. As shown in Fig.~\ref{fig:SIfig_n_w/d} for $\mu=0.35$ and both low ($\mu_r=0.001$) and high ($\mu_r=1$) rolling friction, the average arch size $\langle n \rangle$ scales linearly with $w/d$, consistent with previous experimental observations~\cite{garcimartin_shape_2010}. The slope of this linear dependence exhibits a crossover from roughly $1$ at low $\mu_r$ to roughly $1.8$ at large $\mu_r$ (inset of Fig.~\ref{fig:SIfig_n_w/d}). This demonstrates that particles with blocked rolling form arches comprising more particles than freely rolling ones, consistent with the reported arch size distributions at high and low-$\mu_r$ (Fig.~\textcolor{blue}{3}).

\section{Robustness of minimal cluster size measurements}

\begin{figure}[b!]
\includegraphics[width=1\linewidth]{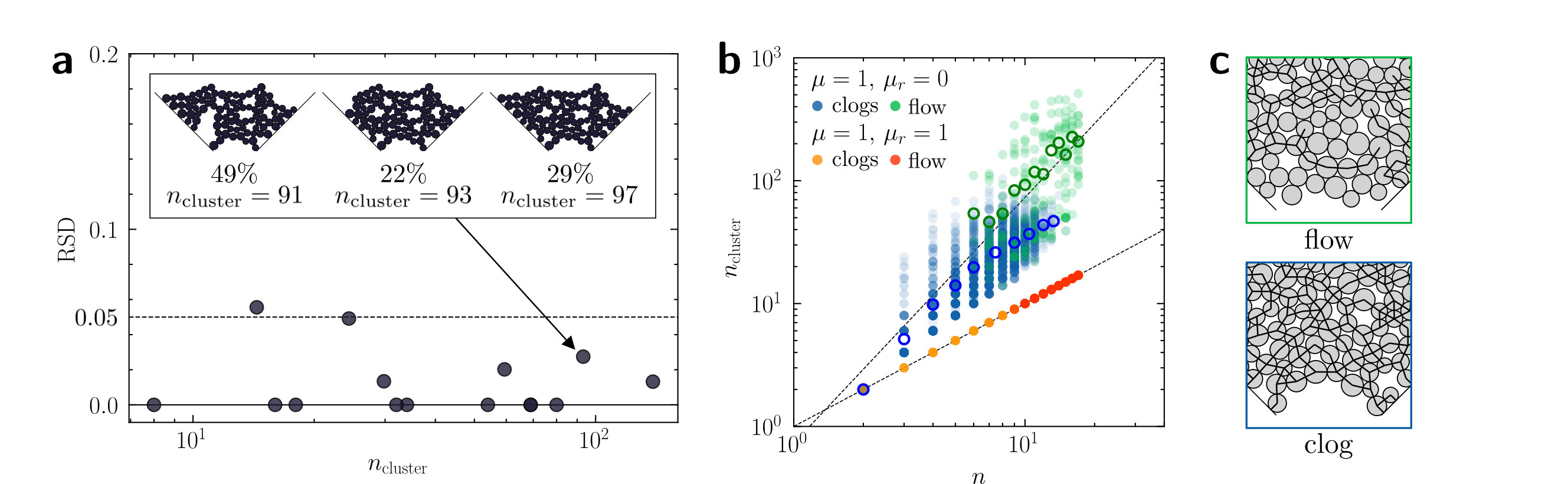}
\caption{\label{fig:SIfig_pg} \textbf{Robustness of minimal cluster size measurements.} \textbf{(a)} Relative standard deviation (RSD) of the minimal cluster size versus the average size, $\langle \nclus \rangle$, for various packing configurations. Each data point is computed from 20 independent trials. The horizontal dashed line indicates a 5\% RSD. Inset: Representative snapshots of distinct minimal clusters obtained from the same initial configuration, corresponding to the data point indicated by the arrow. \textbf{(b)} Minimal cluster size $\nclus$ versus the arch size $n$. Individual measurements are shown for $\mu_r=0$ (blue: clogs, green: flow) and for $\mu_r=1$ (orange: clogs, red: flow). Open symbols show the average cluster size for $\mu_r=0$. Dashed lines indicate the linear and quadratic fits shown in Fig.~\textcolor{blue}{4}(b). \textbf{(c)} Representative snapshots of the contact network during flow (top) and in clogs (bottom), illustrating the lower mean contact number per particle during discharge.}
\end{figure}

Minimal clogging clusters are obtained by iteratively removing randomly selected particles from the system's upper boundary until all remaining particles are needed for arch rigidity. Since this method relies on the random selection of particles, it may result in slightly different minimal clogging clusters depending on the order of removal. To quantify this deviation, we ran the algorithm 20 times on a selected subset of packing configurations with $\mu=1$ and $\mu_r=0$, systematically measuring the average cluster size $\langle \nclus \rangle$ for each sampled configuration. The relative standard deviation (RSD) of these 20 measurements is plotted against the average cluster size in Fig.~\ref{fig:SIfig_pg}(a) alongside representative snapshots of different minimal clusters obtained from a single initial configuration. The maximum RSD remains around $5\%$, with many configurations consistently yielding the exact same value of $\nclus$. This variance is substantially smaller than the standard deviation of $\nclus$ between different packing configurations, confirming the robustness of our measurements.

As detailed in the Methods, minimal clusters are identified in both flowing and clogged configurations to ensure unbiased measurements of $\nclus$ over a wide range of arch sizes $n$. While rigid clusters at small $n$ are only observed in clogged configurations due to their high clogging probability, those at large $n$ are only observed in flowing configurations due to their extremely low probability of flow arrest (Fig.~\ref{fig:SIfig_pg}(b)). At intermediate $n$, however, minimal clusters are detected in both clogged and flowing configurations, leading to a sampling bias. Because larger clusters are more likely to be sampled during flow whereas smaller ones are more likely to form stable clogs, the measured $\nclus$ is underestimated in clogs and overestimated during flow. To avoid this sampling bias, the data shown in Fig.~\textcolor{blue}{4}(b) are restricted to $n\leq8$ in clogged configurations and to $n\geq7$ in flowing ones. For $\mu_r=1$, however, this bias does not occur because spanning arches with blocked sliding and rolling are intrinsically rigid, yielding $\nclus = n$ regardless of whether the system is flowing or clogged.

Clogged and flowing configurations also differ in the connectivity of the contact networks. Because flowing configurations are less densely packed than clogged ones and have a lower mean contact number per particle (Fig.~\ref{fig:SIfig_pg}(c)), spanning arches require more wall contacts to achieve rigidity during flow. As a result, the average cluster size at fixed $n$ is slightly larger in flowing configurations than in clogged ones, as observed in Fig.~\textcolor{blue}{4}(b).

\section{Influence of the hopper angle}

\begin{figure}[h!]
\includegraphics[width=1\linewidth]{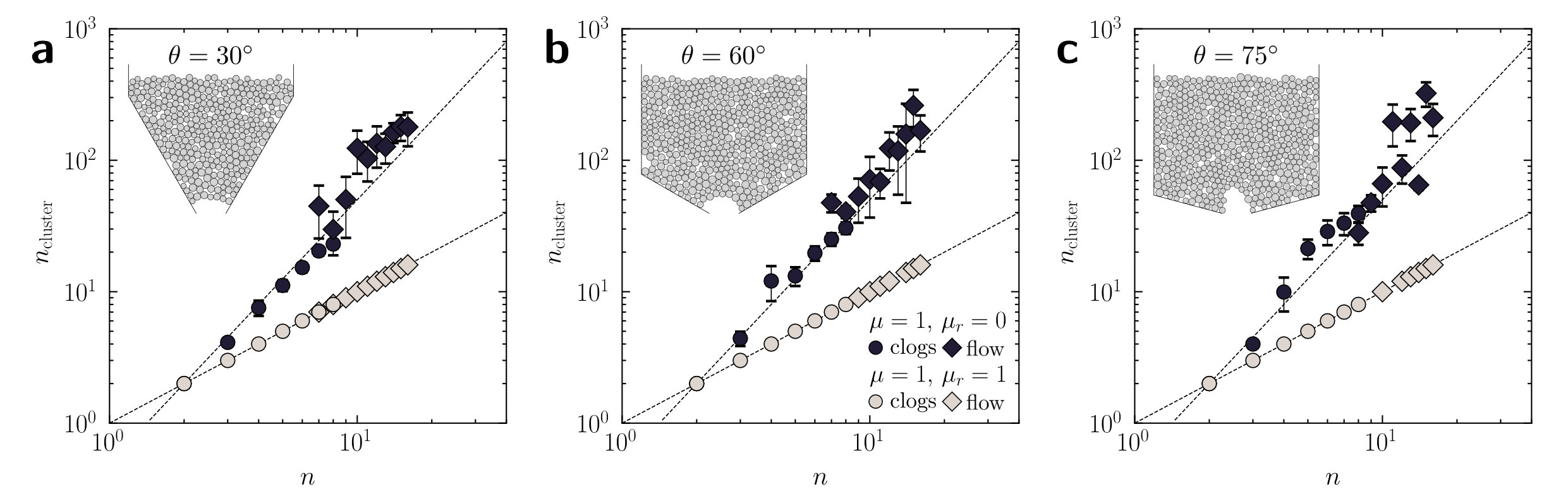}
\caption{\label{fig:SIfig_angle} \textbf{Minimal cluster size for varying hopper angles.} Average cluster size $\nclus$ versus arch size $n$ in hopper channels with angles \textbf{(a)} $\theta=30^{\circ}$, \textbf{(b)} $\theta=60^{\circ}$, and \textbf{(c)} $\theta=75^{\circ}$. Data are shown for frictional particles ($\mu=1$) with free rolling ($\mu_r=0$, dark blue symbols) and blocked rolling ($\mu_r=1$, beige symbols), both in clogs (circles) and during flow (diamonds). Dashed lines correspond to $\nclus = 0.5 n^2$ and $\nclus = n$. Error bars indicate 95\% confidence intervals of the average cluster size within each bin.}
\end{figure}

The main text focuses on hopper channels with a fixed angle $\theta=45^{\circ}$. To verify that the quadratic and linear scalings of the minimal cluster size for particles with free and blocked rolling are independent of the hopper angle, we performed additional molecular dynamics simulations with $\theta \in \{30^{\circ}, 60^{\circ}, 75^{\circ}\}$. The minimal cluster sizes were measured following the exact same procedure described in the methods section. The average cluster size $\nclus$ in clogs and during flow is plotted against the arch size $n$ for each channel geometry in Fig.~\ref{fig:SIfig_angle}. Consistent with the results presented in the main text, the different systems yield the same two scalings, following $\nclus\propto n^2$ for $\mu=1$ and $\mu_r=0$, and $\nclus = n$ for $\mu=1$ and $\mu_r=1$, which suggests the universality of our findings.

\putbib[bibli]
\end{bibunit}


\begin{thebibliography}{0}
\end{thebibliography}

\begin{thebibliography}{49}%
\makeatletter
\providecommand \@ifxundefined [1]{%
 \@ifx{#1\undefined}
}%
\providecommand \@ifnum [1]{%
 \ifnum #1\expandafter \@firstoftwo
 \else \expandafter \@secondoftwo
 \fi
}%
\providecommand \@ifx [1]{%
 \ifx #1\expandafter \@firstoftwo
 \else \expandafter \@secondoftwo
 \fi
}%
\providecommand \natexlab [1]{#1}%
\providecommand \enquote  [1]{``#1''}%
\providecommand \bibnamefont  [1]{#1}%
\providecommand \bibfnamefont [1]{#1}%
\providecommand \citenamefont [1]{#1}%
\providecommand \href@noop [0]{\@secondoftwo}%
\providecommand \href [0]{\begingroup \@sanitize@url \@href}%
\providecommand \@href[1]{\@@startlink{#1}\@@href}%
\providecommand \@@href[1]{\endgroup#1\@@endlink}%
\providecommand \@sanitize@url [0]{\catcode `\\12\catcode `\$12\catcode `\&12\catcode `\#12\catcode `\^12\catcode `\_12\catcode `\%12\relax}%
\providecommand \@@startlink[1]{}%
\providecommand \@@endlink[0]{}%
\providecommand \url  [0]{\begingroup\@sanitize@url \@url }%
\providecommand \@url [1]{\endgroup\@href {#1}{\urlprefix }}%
\providecommand \urlprefix  [0]{URL }%
\providecommand \Eprint [0]{\href }%
\providecommand \doibase [0]{https://doi.org/}%
\providecommand \selectlanguage [0]{\@gobble}%
\providecommand \bibinfo  [0]{\@secondoftwo}%
\providecommand \bibfield  [0]{\@secondoftwo}%
\providecommand \translation [1]{[#1]}%
\providecommand \BibitemOpen [0]{}%
\providecommand \bibitemStop [0]{}%
\providecommand \bibitemNoStop [0]{.\EOS\space}%
\providecommand \EOS [0]{\spacefactor3000\relax}%
\providecommand \BibitemShut  [1]{\csname bibitem#1\endcsname}%
\let\auto@bib@innerbib\@empty
\bibitem [{\citenamefont {Jaeger}\ \emph {et~al.}(1996)\citenamefont {Jaeger}, \citenamefont {Nagel},\ and\ \citenamefont {Behringer}}]{jaeger_granular_1996}%
  \BibitemOpen
  \bibfield  {author} {\bibinfo {author} {\bibfnamefont {H.~M.}\ \bibnamefont {Jaeger}}, \bibinfo {author} {\bibfnamefont {S.~R.}\ \bibnamefont {Nagel}},\ and\ \bibinfo {author} {\bibfnamefont {R.~P.}\ \bibnamefont {Behringer}},\ }\href {https://doi.org/10.1103/RevModPhys.68.1259} {\bibfield  {journal} {\bibinfo  {journal} {Reviews of Modern Physics}\ }\textbf {\bibinfo {volume} {68}},\ \bibinfo {pages} {1259} (\bibinfo {year} {1996})}\BibitemShut {NoStop}%
\bibitem [{\citenamefont {Goyon}\ \emph {et~al.}(2008)\citenamefont {Goyon}, \citenamefont {Colin}, \citenamefont {Ovarlez}, \citenamefont {Ajdari},\ and\ \citenamefont {Bocquet}}]{goyon_spatial_2008}%
  \BibitemOpen
  \bibfield  {author} {\bibinfo {author} {\bibfnamefont {J.}~\bibnamefont {Goyon}}, \bibinfo {author} {\bibfnamefont {A.}~\bibnamefont {Colin}}, \bibinfo {author} {\bibfnamefont {G.}~\bibnamefont {Ovarlez}}, \bibinfo {author} {\bibfnamefont {A.}~\bibnamefont {Ajdari}},\ and\ \bibinfo {author} {\bibfnamefont {L.}~\bibnamefont {Bocquet}},\ }\href {https://doi.org/10.1038/nature07026} {\bibfield  {journal} {\bibinfo  {journal} {Nature}\ }\textbf {\bibinfo {volume} {454}},\ \bibinfo {pages} {84} (\bibinfo {year} {2008})}\BibitemShut {NoStop}%
\bibitem [{\citenamefont {Gu}\ \emph {et~al.}(2025)\citenamefont {Gu}, \citenamefont {Guiselin}, \citenamefont {Bain}, \citenamefont {Zuriguel},\ and\ \citenamefont {Bartolo}}]{gu_emergence_2025}%
  \BibitemOpen
  \bibfield  {author} {\bibinfo {author} {\bibfnamefont {F.}~\bibnamefont {Gu}}, \bibinfo {author} {\bibfnamefont {B.}~\bibnamefont {Guiselin}}, \bibinfo {author} {\bibfnamefont {N.}~\bibnamefont {Bain}}, \bibinfo {author} {\bibfnamefont {I.}~\bibnamefont {Zuriguel}},\ and\ \bibinfo {author} {\bibfnamefont {D.}~\bibnamefont {Bartolo}},\ }\href {https://doi.org/10.1038/s41586-024-08514-6} {\bibfield  {journal} {\bibinfo  {journal} {Nature}\ }\textbf {\bibinfo {volume} {638}},\ \bibinfo {pages} {112} (\bibinfo {year} {2025})}\BibitemShut {NoStop}%
\bibitem [{\citenamefont {Bantawa}\ \emph {et~al.}(2023)\citenamefont {Bantawa}, \citenamefont {Keshavarz}, \citenamefont {Geri}, \citenamefont {Bouzid}, \citenamefont {Divoux}, \citenamefont {McKinley},\ and\ \citenamefont {Del~Gado}}]{bantawa_hidden_2023}%
  \BibitemOpen
  \bibfield  {author} {\bibinfo {author} {\bibfnamefont {M.}~\bibnamefont {Bantawa}}, \bibinfo {author} {\bibfnamefont {B.}~\bibnamefont {Keshavarz}}, \bibinfo {author} {\bibfnamefont {M.}~\bibnamefont {Geri}}, \bibinfo {author} {\bibfnamefont {M.}~\bibnamefont {Bouzid}}, \bibinfo {author} {\bibfnamefont {T.}~\bibnamefont {Divoux}}, \bibinfo {author} {\bibfnamefont {G.~H.}\ \bibnamefont {McKinley}},\ and\ \bibinfo {author} {\bibfnamefont {E.}~\bibnamefont {Del~Gado}},\ }\href {https://doi.org/10.1038/s41567-023-01988-7} {\bibfield  {journal} {\bibinfo  {journal} {Nature Physics}\ }\textbf {\bibinfo {volume} {19}},\ \bibinfo {pages} {1178} (\bibinfo {year} {2023})}\BibitemShut {NoStop}%
\bibitem [{\citenamefont {Bi}\ \emph {et~al.}(2015)\citenamefont {Bi}, \citenamefont {Lopez}, \citenamefont {Schwarz},\ and\ \citenamefont {Manning}}]{bi_density-independent_2015}%
  \BibitemOpen
  \bibfield  {author} {\bibinfo {author} {\bibfnamefont {D.}~\bibnamefont {Bi}}, \bibinfo {author} {\bibfnamefont {J.~H.}\ \bibnamefont {Lopez}}, \bibinfo {author} {\bibfnamefont {J.~M.}\ \bibnamefont {Schwarz}},\ and\ \bibinfo {author} {\bibfnamefont {M.~L.}\ \bibnamefont {Manning}},\ }\href {https://doi.org/10.1038/nphys3471} {\bibfield  {journal} {\bibinfo  {journal} {Nature Physics}\ }\textbf {\bibinfo {volume} {11}},\ \bibinfo {pages} {1074} (\bibinfo {year} {2015})}\BibitemShut {NoStop}%
\bibitem [{\citenamefont {Liu}\ and\ \citenamefont {Nagel}(1998)}]{liu_jamming_1998}%
  \BibitemOpen
  \bibfield  {author} {\bibinfo {author} {\bibfnamefont {A.~J.}\ \bibnamefont {Liu}}\ and\ \bibinfo {author} {\bibfnamefont {S.~R.}\ \bibnamefont {Nagel}},\ }\href {https://doi.org/10.1038/23819} {\bibfield  {journal} {\bibinfo  {journal} {Nature}\ }\textbf {\bibinfo {volume} {396}},\ \bibinfo {pages} {21} (\bibinfo {year} {1998})}\BibitemShut {NoStop}%
\bibitem [{\citenamefont {Dashti}\ \emph {et~al.}(2023)\citenamefont {Dashti}, \citenamefont {Saberi}, \citenamefont {Rahbari},\ and\ \citenamefont {Kurths}}]{dashti_emergence_2023}%
  \BibitemOpen
  \bibfield  {author} {\bibinfo {author} {\bibfnamefont {H.}~\bibnamefont {Dashti}}, \bibinfo {author} {\bibfnamefont {A.~A.}\ \bibnamefont {Saberi}}, \bibinfo {author} {\bibfnamefont {S.}~\bibnamefont {Rahbari}},\ and\ \bibinfo {author} {\bibfnamefont {J.}~\bibnamefont {Kurths}},\ }\href {https://doi.org/10.1126/sciadv.adh5586} {\bibfield  {journal} {\bibinfo  {journal} {Science Advances}\ }\textbf {\bibinfo {volume} {9}},\ \bibinfo {pages} {eadh5586} (\bibinfo {year} {2023})}\BibitemShut {NoStop}%
\bibitem [{\citenamefont {Zuriguel}\ \emph {et~al.}(2014)\citenamefont {Zuriguel}, \citenamefont {Parisi}, \citenamefont {Hidalgo}, \citenamefont {Lozano}, \citenamefont {Janda}, \citenamefont {Gago}, \citenamefont {Peralta}, \citenamefont {Ferrer}, \citenamefont {Pugnaloni}, \citenamefont {Clément}, \citenamefont {Maza}, \citenamefont {Pagonabarraga},\ and\ \citenamefont {Garcimartín}}]{zuriguel_clogging_2014}%
  \BibitemOpen
  \bibfield  {author} {\bibinfo {author} {\bibfnamefont {I.}~\bibnamefont {Zuriguel}}, \bibinfo {author} {\bibfnamefont {D.~R.}\ \bibnamefont {Parisi}}, \bibinfo {author} {\bibfnamefont {R.~C.}\ \bibnamefont {Hidalgo}}, \bibinfo {author} {\bibfnamefont {C.}~\bibnamefont {Lozano}}, \bibinfo {author} {\bibfnamefont {A.}~\bibnamefont {Janda}}, \bibinfo {author} {\bibfnamefont {P.~A.}\ \bibnamefont {Gago}}, \bibinfo {author} {\bibfnamefont {J.~P.}\ \bibnamefont {Peralta}}, \bibinfo {author} {\bibfnamefont {L.~M.}\ \bibnamefont {Ferrer}}, \bibinfo {author} {\bibfnamefont {L.~A.}\ \bibnamefont {Pugnaloni}}, \bibinfo {author} {\bibfnamefont {E.}~\bibnamefont {Clément}}, \bibinfo {author} {\bibfnamefont {D.}~\bibnamefont {Maza}}, \bibinfo {author} {\bibfnamefont {I.}~\bibnamefont {Pagonabarraga}},\ and\ \bibinfo {author} {\bibfnamefont {A.}~\bibnamefont {Garcimartín}},\ }\href {https://doi.org/10.1038/srep07324} {\bibfield  {journal} {\bibinfo  {journal} {Scientific Reports}\ }\textbf {\bibinfo {volume} {4}},\
  \bibinfo {pages} {7324} (\bibinfo {year} {2014})}\BibitemShut {NoStop}%
\bibitem [{\citenamefont {Dressaire}\ and\ \citenamefont {Sauret}(2016)}]{dressaire_clogging_2016}%
  \BibitemOpen
  \bibfield  {author} {\bibinfo {author} {\bibfnamefont {E.}~\bibnamefont {Dressaire}}\ and\ \bibinfo {author} {\bibfnamefont {A.}~\bibnamefont {Sauret}},\ }\href {https://doi.org/10.1039/C6SM01879C} {\bibfield  {journal} {\bibinfo  {journal} {Soft Matter}\ }\textbf {\bibinfo {volume} {13}},\ \bibinfo {pages} {37} (\bibinfo {year} {2016})}\BibitemShut {NoStop}%
\bibitem [{\citenamefont {Marin}\ and\ \citenamefont {Souzy}(2025)}]{marin_clogging_2025}%
  \BibitemOpen
  \bibfield  {author} {\bibinfo {author} {\bibfnamefont {A.}~\bibnamefont {Marin}}\ and\ \bibinfo {author} {\bibfnamefont {M.}~\bibnamefont {Souzy}},\ }\href {https://doi.org/10.1146/annurev-fluid-030124-112742} {\bibfield  {journal} {\bibinfo  {journal} {Annual Review of Fluid Mechanics}\ }\textbf {\bibinfo {volume} {57}},\ \bibinfo {pages} {89} (\bibinfo {year} {2025})}\BibitemShut {NoStop}%
\bibitem [{\citenamefont {Schulze}(2021)}]{schulze_powders_2021}%
  \BibitemOpen
  \bibfield  {author} {\bibinfo {author} {\bibfnamefont {D.}~\bibnamefont {Schulze}},\ }\href@noop {} {{\selectlanguage {en}\emph {\bibinfo {title} {Powders and {Bulk} {Solids}: {Behavior}, {Characterization}, {Storage} and {Flow}}}}}\ (\bibinfo  {publisher} {Springer Nature},\ \bibinfo {year} {2021})\BibitemShut {NoStop}%
\bibitem [{\citenamefont {Bakhtiari}\ and\ \citenamefont {Kähler}(2024)}]{bakhtiari_method_2024}%
  \BibitemOpen
  \bibfield  {author} {\bibinfo {author} {\bibfnamefont {A.}~\bibnamefont {Bakhtiari}}\ and\ \bibinfo {author} {\bibfnamefont {C.~J.}\ \bibnamefont {Kähler}},\ }\href {https://doi.org/10.1063/5.0214436} {\bibfield  {journal} {\bibinfo  {journal} {Biomicrofluidics}\ }\textbf {\bibinfo {volume} {18}},\ \bibinfo {pages} {044101} (\bibinfo {year} {2024})}\BibitemShut {NoStop}%
\bibitem [{\citenamefont {Shelby}\ \emph {et~al.}(2003)\citenamefont {Shelby}, \citenamefont {White}, \citenamefont {Ganesan}, \citenamefont {Rathod},\ and\ \citenamefont {Chiu}}]{shelby_microfluidic_2003}%
  \BibitemOpen
  \bibfield  {author} {\bibinfo {author} {\bibfnamefont {J.~P.}\ \bibnamefont {Shelby}}, \bibinfo {author} {\bibfnamefont {J.}~\bibnamefont {White}}, \bibinfo {author} {\bibfnamefont {K.}~\bibnamefont {Ganesan}}, \bibinfo {author} {\bibfnamefont {P.~K.}\ \bibnamefont {Rathod}},\ and\ \bibinfo {author} {\bibfnamefont {D.~T.}\ \bibnamefont {Chiu}},\ }\href {https://doi.org/10.1073/pnas.2433968100} {\bibfield  {journal} {\bibinfo  {journal} {Proceedings of the National Academy of Sciences}\ }\textbf {\bibinfo {volume} {100}},\ \bibinfo {pages} {14618} (\bibinfo {year} {2003})}\BibitemShut {NoStop}%
\bibitem [{\citenamefont {Piton}\ \emph {et~al.}(2022)\citenamefont {Piton}, \citenamefont {Goodwin}, \citenamefont {Mark},\ and\ \citenamefont {Strouth}}]{piton_debris_2022}%
  \BibitemOpen
  \bibfield  {author} {\bibinfo {author} {\bibfnamefont {G.}~\bibnamefont {Piton}}, \bibinfo {author} {\bibfnamefont {S.~R.}\ \bibnamefont {Goodwin}}, \bibinfo {author} {\bibfnamefont {E.}~\bibnamefont {Mark}},\ and\ \bibinfo {author} {\bibfnamefont {A.}~\bibnamefont {Strouth}},\ }\href {https://doi.org/10.1029/2021JF006447} {\bibfield  {journal} {\bibinfo  {journal} {Journal of Geophysical Research: Earth Surface}\ }\textbf {\bibinfo {volume} {127}},\ \bibinfo {pages} {e2021JF006447} (\bibinfo {year} {2022})}\BibitemShut {NoStop}%
\bibitem [{\citenamefont {Clément}\ \emph {et~al.}(2000)\citenamefont {Clément}, \citenamefont {Reydellet}, \citenamefont {Rioual}, \citenamefont {Parise}, \citenamefont {Fanguet}, \citenamefont {Lanuza},\ and\ \citenamefont {Kolb}}]{clement_jamming_2000}%
  \BibitemOpen
  \bibfield  {author} {\bibinfo {author} {\bibfnamefont {E.}~\bibnamefont {Clément}}, \bibinfo {author} {\bibfnamefont {G.}~\bibnamefont {Reydellet}}, \bibinfo {author} {\bibfnamefont {F.}~\bibnamefont {Rioual}}, \bibinfo {author} {\bibfnamefont {B.}~\bibnamefont {Parise}}, \bibinfo {author} {\bibfnamefont {V.}~\bibnamefont {Fanguet}}, \bibinfo {author} {\bibfnamefont {J.}~\bibnamefont {Lanuza}},\ and\ \bibinfo {author} {\bibfnamefont {E.}~\bibnamefont {Kolb}},\ }in\ \href {https://doi.org/10.1007/978-3-642-59751-0_50} {{\selectlanguage {en}\emph {\bibinfo {booktitle} {Traffic and {Granular} {Flow} ’99}}}},\ \bibinfo {editor} {edited by\ \bibinfo {editor} {\bibfnamefont {D.}~\bibnamefont {Helbing}}, \bibinfo {editor} {\bibfnamefont {H.~J.}\ \bibnamefont {Herrmann}}, \bibinfo {editor} {\bibfnamefont {M.}~\bibnamefont {Schreckenberg}},\ and\ \bibinfo {editor} {\bibfnamefont {D.~E.}\ \bibnamefont {Wolf}}}\ (\bibinfo  {publisher} {Springer},\ \bibinfo {address} {Berlin, Heidelberg},\ \bibinfo {year} {2000})\
  pp.\ \bibinfo {pages} {457--468}\BibitemShut {NoStop}%
\bibitem [{\citenamefont {Zuriguel}\ \emph {et~al.}(2005)\citenamefont {Zuriguel}, \citenamefont {Garcimartín}, \citenamefont {Maza}, \citenamefont {Pugnaloni},\ and\ \citenamefont {Pastor}}]{zuriguel_jamming_2005}%
  \BibitemOpen
  \bibfield  {author} {\bibinfo {author} {\bibfnamefont {I.}~\bibnamefont {Zuriguel}}, \bibinfo {author} {\bibfnamefont {A.}~\bibnamefont {Garcimartín}}, \bibinfo {author} {\bibfnamefont {D.}~\bibnamefont {Maza}}, \bibinfo {author} {\bibfnamefont {L.~A.}\ \bibnamefont {Pugnaloni}},\ and\ \bibinfo {author} {\bibfnamefont {J.~M.}\ \bibnamefont {Pastor}},\ }\href {https://doi.org/10.1103/PhysRevE.71.051303} {\bibfield  {journal} {\bibinfo  {journal} {Physical Review E}\ }\textbf {\bibinfo {volume} {71}},\ \bibinfo {pages} {051303} (\bibinfo {year} {2005})}\BibitemShut {NoStop}%
\bibitem [{\citenamefont {Janda}\ \emph {et~al.}(2008)\citenamefont {Janda}, \citenamefont {Zuriguel}, \citenamefont {Garcimartín}, \citenamefont {Pugnaloni},\ and\ \citenamefont {Maza}}]{janda_jamming_2008}%
  \BibitemOpen
  \bibfield  {author} {\bibinfo {author} {\bibfnamefont {A.}~\bibnamefont {Janda}}, \bibinfo {author} {\bibfnamefont {I.}~\bibnamefont {Zuriguel}}, \bibinfo {author} {\bibfnamefont {A.}~\bibnamefont {Garcimartín}}, \bibinfo {author} {\bibfnamefont {L.~A.}\ \bibnamefont {Pugnaloni}},\ and\ \bibinfo {author} {\bibfnamefont {D.}~\bibnamefont {Maza}},\ }\href {https://doi.org/10.1209/0295-5075/84/44002} {\bibfield  {journal} {\bibinfo  {journal} {Europhysics Letters}\ }\textbf {\bibinfo {volume} {84}},\ \bibinfo {pages} {44002} (\bibinfo {year} {2008})}\BibitemShut {NoStop}%
\bibitem [{\citenamefont {Hong}\ \emph {et~al.}(2017)\citenamefont {Hong}, \citenamefont {Kohne}, \citenamefont {Morrell}, \citenamefont {Wang},\ and\ \citenamefont {Weeks}}]{hong_clogging_2017}%
  \BibitemOpen
  \bibfield  {author} {\bibinfo {author} {\bibfnamefont {X.}~\bibnamefont {Hong}}, \bibinfo {author} {\bibfnamefont {M.}~\bibnamefont {Kohne}}, \bibinfo {author} {\bibfnamefont {M.}~\bibnamefont {Morrell}}, \bibinfo {author} {\bibfnamefont {H.}~\bibnamefont {Wang}},\ and\ \bibinfo {author} {\bibfnamefont {E.~R.}\ \bibnamefont {Weeks}},\ }\href {https://doi.org/10.1103/PhysRevE.96.062605} {\bibfield  {journal} {\bibinfo  {journal} {Physical Review E}\ }\textbf {\bibinfo {volume} {96}},\ \bibinfo {pages} {062605} (\bibinfo {year} {2017})}\BibitemShut {NoStop}%
\bibitem [{\citenamefont {Hsu}\ \emph {et~al.}(2021)\citenamefont {Hsu}, \citenamefont {Baysal}, \citenamefont {Wirenborn}, \citenamefont {Mårtensson}, \citenamefont {Wittberg},\ and\ \citenamefont {Isa}}]{hsu_roughness-dependent_2021}%
  \BibitemOpen
  \bibfield  {author} {\bibinfo {author} {\bibfnamefont {C.-P.}\ \bibnamefont {Hsu}}, \bibinfo {author} {\bibfnamefont {H.~E.}\ \bibnamefont {Baysal}}, \bibinfo {author} {\bibfnamefont {G.}~\bibnamefont {Wirenborn}}, \bibinfo {author} {\bibfnamefont {G.}~\bibnamefont {Mårtensson}}, \bibinfo {author} {\bibfnamefont {L.~P.}\ \bibnamefont {Wittberg}},\ and\ \bibinfo {author} {\bibfnamefont {L.}~\bibnamefont {Isa}},\ }\href {https://doi.org/10.1039/D1SM00738F} {\bibfield  {journal} {\bibinfo  {journal} {Soft Matter}\ }\textbf {\bibinfo {volume} {17}},\ \bibinfo {pages} {7252} (\bibinfo {year} {2021})}\BibitemShut {NoStop}%
\bibitem [{\citenamefont {Vani}\ \emph {et~al.}(2022)\citenamefont {Vani}, \citenamefont {Escudier},\ and\ \citenamefont {Sauret}}]{vani_influence_2022}%
  \BibitemOpen
  \bibfield  {author} {\bibinfo {author} {\bibfnamefont {N.}~\bibnamefont {Vani}}, \bibinfo {author} {\bibfnamefont {S.}~\bibnamefont {Escudier}},\ and\ \bibinfo {author} {\bibfnamefont {A.}~\bibnamefont {Sauret}},\ }\href {https://doi.org/10.1039/D2SM00962E} {\bibfield  {journal} {\bibinfo  {journal} {Soft Matter}\ }\textbf {\bibinfo {volume} {18}},\ \bibinfo {pages} {6987} (\bibinfo {year} {2022})}\BibitemShut {NoStop}%
\bibitem [{\citenamefont {Hoggarth}\ \emph {et~al.}(2026)\citenamefont {Hoggarth}, \citenamefont {Illing}, \citenamefont {Weeks},\ and\ \citenamefont {Dalnoki-Veress}}]{hoggarth_clogging_2026}%
  \BibitemOpen
  \bibfield  {author} {\bibinfo {author} {\bibfnamefont {J.}~\bibnamefont {Hoggarth}}, \bibinfo {author} {\bibfnamefont {P.~E.}\ \bibnamefont {Illing}}, \bibinfo {author} {\bibfnamefont {E.~R.}\ \bibnamefont {Weeks}},\ and\ \bibinfo {author} {\bibfnamefont {K.}~\bibnamefont {Dalnoki-Veress}},\ }\href {https://doi.org/10.1103/n68l-9b38} {\bibfield  {journal} {\bibinfo  {journal} {Physical Review Letters}\ }\textbf {\bibinfo {volume} {136}},\ \bibinfo {pages} {088201} (\bibinfo {year} {2026})}\BibitemShut {NoStop}%
\bibitem [{\citenamefont {Zuriguel}\ \emph {et~al.}(2011)\citenamefont {Zuriguel}, \citenamefont {Janda}, \citenamefont {Garcimartín}, \citenamefont {Lozano}, \citenamefont {Arévalo},\ and\ \citenamefont {Maza}}]{zuriguel_silo_2011}%
  \BibitemOpen
  \bibfield  {author} {\bibinfo {author} {\bibfnamefont {I.}~\bibnamefont {Zuriguel}}, \bibinfo {author} {\bibfnamefont {A.}~\bibnamefont {Janda}}, \bibinfo {author} {\bibfnamefont {A.}~\bibnamefont {Garcimartín}}, \bibinfo {author} {\bibfnamefont {C.}~\bibnamefont {Lozano}}, \bibinfo {author} {\bibfnamefont {R.}~\bibnamefont {Arévalo}},\ and\ \bibinfo {author} {\bibfnamefont {D.}~\bibnamefont {Maza}},\ }\href {https://doi.org/10.1103/PhysRevLett.107.278001} {\bibfield  {journal} {\bibinfo  {journal} {Physical Review Letters}\ }\textbf {\bibinfo {volume} {107}},\ \bibinfo {pages} {278001} (\bibinfo {year} {2011})}\BibitemShut {NoStop}%
\bibitem [{\citenamefont {López-Rodríguez}\ \emph {et~al.}(2019)\citenamefont {López-Rodríguez}, \citenamefont {Gella}, \citenamefont {To}, \citenamefont {Maza}, \citenamefont {Garcimartín},\ and\ \citenamefont {Zuriguel}}]{lopez-rodriguez_effect_2019}%
  \BibitemOpen
  \bibfield  {author} {\bibinfo {author} {\bibfnamefont {D.}~\bibnamefont {López-Rodríguez}}, \bibinfo {author} {\bibfnamefont {D.}~\bibnamefont {Gella}}, \bibinfo {author} {\bibfnamefont {K.}~\bibnamefont {To}}, \bibinfo {author} {\bibfnamefont {D.}~\bibnamefont {Maza}}, \bibinfo {author} {\bibfnamefont {A.}~\bibnamefont {Garcimartín}},\ and\ \bibinfo {author} {\bibfnamefont {I.}~\bibnamefont {Zuriguel}},\ }\href {https://doi.org/10.1103/PhysRevE.99.032901} {\bibfield  {journal} {\bibinfo  {journal} {Physical Review. E}\ }\textbf {\bibinfo {volume} {99}},\ \bibinfo {pages} {032901} (\bibinfo {year} {2019})}\BibitemShut {NoStop}%
\bibitem [{\citenamefont {Vani}\ \emph {et~al.}(2024)\citenamefont {Vani}, \citenamefont {Escudier}, \citenamefont {Jeong},\ and\ \citenamefont {Sauret}}]{vani_role_2024}%
  \BibitemOpen
  \bibfield  {author} {\bibinfo {author} {\bibfnamefont {N.}~\bibnamefont {Vani}}, \bibinfo {author} {\bibfnamefont {S.}~\bibnamefont {Escudier}}, \bibinfo {author} {\bibfnamefont {D.-H.}\ \bibnamefont {Jeong}},\ and\ \bibinfo {author} {\bibfnamefont {A.}~\bibnamefont {Sauret}},\ }\href {https://doi.org/10.1103/PhysRevResearch.6.L032060} {\bibfield  {journal} {\bibinfo  {journal} {Physical Review Research}\ }\textbf {\bibinfo {volume} {6}},\ \bibinfo {pages} {L032060} (\bibinfo {year} {2024})}\BibitemShut {NoStop}%
\bibitem [{\citenamefont {To}(2005)}]{to_jamming_2005}%
  \BibitemOpen
  \bibfield  {author} {\bibinfo {author} {\bibfnamefont {K.}~\bibnamefont {To}},\ }\href {https://doi.org/10.1103/PhysRevE.71.060301} {\bibfield  {journal} {\bibinfo  {journal} {Physical Review E}\ }\textbf {\bibinfo {volume} {71}},\ \bibinfo {pages} {060301} (\bibinfo {year} {2005})}\BibitemShut {NoStop}%
\bibitem [{\citenamefont {Thomas}\ and\ \citenamefont {Durian}(2015)}]{thomas_fraction_2015}%
  \BibitemOpen
  \bibfield  {author} {\bibinfo {author} {\bibfnamefont {C.~C.}\ \bibnamefont {Thomas}}\ and\ \bibinfo {author} {\bibfnamefont {D.~J.}\ \bibnamefont {Durian}},\ }\href {https://doi.org/10.1103/PhysRevLett.114.178001} {\bibfield  {journal} {\bibinfo  {journal} {Physical Review Letters}\ }\textbf {\bibinfo {volume} {114}},\ \bibinfo {pages} {178001} (\bibinfo {year} {2015})}\BibitemShut {NoStop}%
\bibitem [{\citenamefont {Montero}\ \emph {et~al.}(2025)\citenamefont {Montero}, \citenamefont {Pugnaloni},\ and\ \citenamefont {Kozlowski}}]{montero_dimensional_2025}%
  \BibitemOpen
  \bibfield  {author} {\bibinfo {author} {\bibfnamefont {J.}~\bibnamefont {Montero}}, \bibinfo {author} {\bibfnamefont {L.~A.}\ \bibnamefont {Pugnaloni}},\ and\ \bibinfo {author} {\bibfnamefont {R.}~\bibnamefont {Kozlowski}},\ }\href {https://doi.org/10.1103/g9yk-7pf1} {\bibfield  {journal} {\bibinfo  {journal} {Physical Review E}\ }\textbf {\bibinfo {volume} {112}},\ \bibinfo {pages} {065411} (\bibinfo {year} {2025})}\BibitemShut {NoStop}%
\bibitem [{\citenamefont {Goldberg}\ \emph {et~al.}(2018)\citenamefont {Goldberg}, \citenamefont {Carlevaro},\ and\ \citenamefont {Pugnaloni}}]{goldberg_clogging_2018}%
  \BibitemOpen
  \bibfield  {author} {\bibinfo {author} {\bibfnamefont {E.}~\bibnamefont {Goldberg}}, \bibinfo {author} {\bibfnamefont {C.~M.}\ \bibnamefont {Carlevaro}},\ and\ \bibinfo {author} {\bibfnamefont {L.~A.}\ \bibnamefont {Pugnaloni}},\ }\href {https://doi.org/10.1088/1742-5468/aae84b} {\bibfield  {journal} {\bibinfo  {journal} {Journal of Statistical Mechanics: Theory and Experiment}\ }\textbf {\bibinfo {volume} {2018}},\ \bibinfo {pages} {113201} (\bibinfo {year} {2018})}\BibitemShut {NoStop}%
\bibitem [{\citenamefont {Ashour}\ \emph {et~al.}(2017)\citenamefont {Ashour}, \citenamefont {Wegner}, \citenamefont {Trittel}, \citenamefont {Börzsönyi},\ and\ \citenamefont {Stannarius}}]{ashour_outflow_2017}%
  \BibitemOpen
  \bibfield  {author} {\bibinfo {author} {\bibfnamefont {A.}~\bibnamefont {Ashour}}, \bibinfo {author} {\bibfnamefont {S.}~\bibnamefont {Wegner}}, \bibinfo {author} {\bibfnamefont {T.}~\bibnamefont {Trittel}}, \bibinfo {author} {\bibfnamefont {T.}~\bibnamefont {Börzsönyi}},\ and\ \bibinfo {author} {\bibfnamefont {R.}~\bibnamefont {Stannarius}},\ }\href {https://doi.org/10.1039/C6SM02374F} {\bibfield  {journal} {\bibinfo  {journal} {Soft Matter}\ }\textbf {\bibinfo {volume} {13}},\ \bibinfo {pages} {402} (\bibinfo {year} {2017})}\BibitemShut {NoStop}%
\bibitem [{\citenamefont {Gella}\ \emph {et~al.}(2018)\citenamefont {Gella}, \citenamefont {Zuriguel},\ and\ \citenamefont {Maza}}]{gella_decoupling_2018}%
  \BibitemOpen
  \bibfield  {author} {\bibinfo {author} {\bibfnamefont {D.}~\bibnamefont {Gella}}, \bibinfo {author} {\bibfnamefont {I.}~\bibnamefont {Zuriguel}},\ and\ \bibinfo {author} {\bibfnamefont {D.}~\bibnamefont {Maza}},\ }\href {https://doi.org/10.1103/PhysRevLett.121.138001} {\bibfield  {journal} {\bibinfo  {journal} {Physical Review Letters}\ }\textbf {\bibinfo {volume} {121}},\ \bibinfo {pages} {138001} (\bibinfo {year} {2018})}\BibitemShut {NoStop}%
\bibitem [{\citenamefont {Kool}\ \emph {et~al.}(2025)\citenamefont {Kool}, \citenamefont {Tampier}, \citenamefont {Bourrianne},\ and\ \citenamefont {Lindner}}]{kool_hopper_2025}%
  \BibitemOpen
  \bibfield  {author} {\bibinfo {author} {\bibfnamefont {L.}~\bibnamefont {Kool}}, \bibinfo {author} {\bibfnamefont {J.}~\bibnamefont {Tampier}}, \bibinfo {author} {\bibfnamefont {P.}~\bibnamefont {Bourrianne}},\ and\ \bibinfo {author} {\bibfnamefont {A.}~\bibnamefont {Lindner}},\ }\href {https://doi.org/10.1017/flo.2025.10019} {\bibfield  {journal} {\bibinfo  {journal} {Flow}\ }\textbf {\bibinfo {volume} {5}},\ \bibinfo {pages} {E26} (\bibinfo {year} {2025})}\BibitemShut {NoStop}%
\bibitem [{\citenamefont {Hafez}\ \emph {et~al.}(2021)\citenamefont {Hafez}, \citenamefont {Liu}, \citenamefont {Finkbeiner}, \citenamefont {Alouhali}, \citenamefont {Moellendick},\ and\ \citenamefont {Santamarina}}]{hafez_effect_2021}%
  \BibitemOpen
  \bibfield  {author} {\bibinfo {author} {\bibfnamefont {A.}~\bibnamefont {Hafez}}, \bibinfo {author} {\bibfnamefont {Q.}~\bibnamefont {Liu}}, \bibinfo {author} {\bibfnamefont {T.}~\bibnamefont {Finkbeiner}}, \bibinfo {author} {\bibfnamefont {R.~A.}\ \bibnamefont {Alouhali}}, \bibinfo {author} {\bibfnamefont {T.~E.}\ \bibnamefont {Moellendick}},\ and\ \bibinfo {author} {\bibfnamefont {J.~C.}\ \bibnamefont {Santamarina}},\ }\href {https://doi.org/10.1038/s41598-021-82744-w} {\bibfield  {journal} {\bibinfo  {journal} {Scientific Reports}\ }\textbf {\bibinfo {volume} {11}},\ \bibinfo {pages} {3309} (\bibinfo {year} {2021})}\BibitemShut {NoStop}%
\bibitem [{\citenamefont {Saint-Cyr}\ \emph {et~al.}(2011)\citenamefont {Saint-Cyr}, \citenamefont {Delenne}, \citenamefont {Voivret}, \citenamefont {Radjai},\ and\ \citenamefont {Sornay}}]{saint-cyr_rheology_2011}%
  \BibitemOpen
  \bibfield  {author} {\bibinfo {author} {\bibfnamefont {B.}~\bibnamefont {Saint-Cyr}}, \bibinfo {author} {\bibfnamefont {J.-Y.}\ \bibnamefont {Delenne}}, \bibinfo {author} {\bibfnamefont {C.}~\bibnamefont {Voivret}}, \bibinfo {author} {\bibfnamefont {F.}~\bibnamefont {Radjai}},\ and\ \bibinfo {author} {\bibfnamefont {P.}~\bibnamefont {Sornay}},\ }\href {https://doi.org/10.1103/PhysRevE.84.041302} {\bibfield  {journal} {\bibinfo  {journal} {Physical Review E}\ }\textbf {\bibinfo {volume} {84}},\ \bibinfo {pages} {041302} (\bibinfo {year} {2011})}\BibitemShut {NoStop}%
\bibitem [{\citenamefont {Azéma}\ \emph {et~al.}(2013)\citenamefont {Azéma}, \citenamefont {Radjaï}, \citenamefont {Saint-Cyr}, \citenamefont {Delenne},\ and\ \citenamefont {Sornay}}]{azema_rheology_2013}%
  \BibitemOpen
  \bibfield  {author} {\bibinfo {author} {\bibfnamefont {E.}~\bibnamefont {Azéma}}, \bibinfo {author} {\bibfnamefont {F.}~\bibnamefont {Radjaï}}, \bibinfo {author} {\bibfnamefont {B.}~\bibnamefont {Saint-Cyr}}, \bibinfo {author} {\bibfnamefont {J.-Y.}\ \bibnamefont {Delenne}},\ and\ \bibinfo {author} {\bibfnamefont {P.}~\bibnamefont {Sornay}},\ }\href {https://doi.org/10.1103/PhysRevE.87.052205} {\bibfield  {journal} {\bibinfo  {journal} {Physical Review E}\ }\textbf {\bibinfo {volume} {87}},\ \bibinfo {pages} {052205} (\bibinfo {year} {2013})}\BibitemShut {NoStop}%
\bibitem [{\citenamefont {Ai}\ \emph {et~al.}(2011)\citenamefont {Ai}, \citenamefont {Chen}, \citenamefont {Rotter},\ and\ \citenamefont {Ooi}}]{ai_assessment_2011}%
  \BibitemOpen
  \bibfield  {author} {\bibinfo {author} {\bibfnamefont {J.}~\bibnamefont {Ai}}, \bibinfo {author} {\bibfnamefont {J.-F.}\ \bibnamefont {Chen}}, \bibinfo {author} {\bibfnamefont {J.~M.}\ \bibnamefont {Rotter}},\ and\ \bibinfo {author} {\bibfnamefont {J.~Y.}\ \bibnamefont {Ooi}},\ }\href {https://doi.org/10.1016/j.powtec.2010.09.030} {\bibfield  {journal} {\bibinfo  {journal} {Powder Technology}\ }\textbf {\bibinfo {volume} {206}},\ \bibinfo {pages} {269} (\bibinfo {year} {2011})}\BibitemShut {NoStop}%
\bibitem [{\citenamefont {Estrada}\ \emph {et~al.}(2011)\citenamefont {Estrada}, \citenamefont {Azéma}, \citenamefont {Radjai},\ and\ \citenamefont {Taboada}}]{estrada_identification_2011}%
  \BibitemOpen
  \bibfield  {author} {\bibinfo {author} {\bibfnamefont {N.}~\bibnamefont {Estrada}}, \bibinfo {author} {\bibfnamefont {E.}~\bibnamefont {Azéma}}, \bibinfo {author} {\bibfnamefont {F.}~\bibnamefont {Radjai}},\ and\ \bibinfo {author} {\bibfnamefont {A.}~\bibnamefont {Taboada}},\ }\href {https://doi.org/10.1103/PhysRevE.84.011306} {\bibfield  {journal} {\bibinfo  {journal} {Physical Review E}\ }\textbf {\bibinfo {volume} {84}},\ \bibinfo {pages} {011306} (\bibinfo {year} {2011})}\BibitemShut {NoStop}%
\bibitem [{\citenamefont {Singh}\ \emph {et~al.}(2020)\citenamefont {Singh}, \citenamefont {Ness}, \citenamefont {Seto}, \citenamefont {de~Pablo},\ and\ \citenamefont {Jaeger}}]{singh_shear_2020}%
  \BibitemOpen
  \bibfield  {author} {\bibinfo {author} {\bibfnamefont {A.}~\bibnamefont {Singh}}, \bibinfo {author} {\bibfnamefont {C.}~\bibnamefont {Ness}}, \bibinfo {author} {\bibfnamefont {R.}~\bibnamefont {Seto}}, \bibinfo {author} {\bibfnamefont {J.~J.}\ \bibnamefont {de~Pablo}},\ and\ \bibinfo {author} {\bibfnamefont {H.~M.}\ \bibnamefont {Jaeger}},\ }\href {https://doi.org/10.1103/PhysRevLett.124.248005} {\bibfield  {journal} {\bibinfo  {journal} {Physical Review Letters}\ }\textbf {\bibinfo {volume} {124}},\ \bibinfo {pages} {248005} (\bibinfo {year} {2020})}\BibitemShut {NoStop}%
\bibitem [{\citenamefont {Luding}(2008)}]{luding_cohesive_2008}%
  \BibitemOpen
  \bibfield  {author} {\bibinfo {author} {\bibfnamefont {S.}~\bibnamefont {Luding}},\ }\href {https://doi.org/10.1007/s10035-008-0099-x} {\bibfield  {journal} {\bibinfo  {journal} {Granular Matter}\ }\textbf {\bibinfo {volume} {10}},\ \bibinfo {pages} {235} (\bibinfo {year} {2008})}\BibitemShut {NoStop}%
\bibitem [{\citenamefont {To}\ \emph {et~al.}(2001)\citenamefont {To}, \citenamefont {Lai},\ and\ \citenamefont {Pak}}]{to_jamming_2001}%
  \BibitemOpen
  \bibfield  {author} {\bibinfo {author} {\bibfnamefont {K.}~\bibnamefont {To}}, \bibinfo {author} {\bibfnamefont {P.-Y.}\ \bibnamefont {Lai}},\ and\ \bibinfo {author} {\bibfnamefont {H.~K.}\ \bibnamefont {Pak}},\ }\href {https://doi.org/10.1103/PhysRevLett.86.71} {\bibfield  {journal} {\bibinfo  {journal} {Physical Review Letters}\ }\textbf {\bibinfo {volume} {86}},\ \bibinfo {pages} {71} (\bibinfo {year} {2001})}\BibitemShut {NoStop}%
\bibitem [{\citenamefont {Maxwell}(1870)}]{maxwell_ireciprocal_1870}%
  \BibitemOpen
  \bibfield  {author} {\bibinfo {author} {\bibfnamefont {J.~C.}\ \bibnamefont {Maxwell}},\ }\href {https://doi.org/10.1017/S0080456800026351} {\bibfield  {journal} {\bibinfo  {journal} {Earth and Environmental Science Transactions of The Royal Society of Edinburgh}\ }\textbf {\bibinfo {volume} {26}},\ \bibinfo {pages} {1} (\bibinfo {year} {1870})}\BibitemShut {NoStop}%
\bibitem [{\citenamefont {Calladine}(1978)}]{calladine_buckminster_1978}%
  \BibitemOpen
  \bibfield  {author} {\bibinfo {author} {\bibfnamefont {C.~R.}\ \bibnamefont {Calladine}},\ }\href {https://doi.org/10.1016/0020-7683(78)90052-5} {\bibfield  {journal} {\bibinfo  {journal} {International Journal of Solids and Structures}\ }\textbf {\bibinfo {volume} {14}},\ \bibinfo {pages} {161} (\bibinfo {year} {1978})}\BibitemShut {NoStop}%
\bibitem [{\citenamefont {Shundyak}\ \emph {et~al.}(2007)\citenamefont {Shundyak}, \citenamefont {van Hecke},\ and\ \citenamefont {van Saarloos}}]{shundyak_force_2007}%
  \BibitemOpen
  \bibfield  {author} {\bibinfo {author} {\bibfnamefont {K.}~\bibnamefont {Shundyak}}, \bibinfo {author} {\bibfnamefont {M.}~\bibnamefont {van Hecke}},\ and\ \bibinfo {author} {\bibfnamefont {W.}~\bibnamefont {van Saarloos}},\ }\href {https://doi.org/10.1103/PhysRevE.75.010301} {\bibfield  {journal} {\bibinfo  {journal} {Physical Review E}\ }\textbf {\bibinfo {volume} {75}},\ \bibinfo {pages} {010301} (\bibinfo {year} {2007})}\BibitemShut {NoStop}%
\bibitem [{\citenamefont {Jacobs}\ and\ \citenamefont {Thorpe}(1995)}]{jacobs_generic_1995}%
  \BibitemOpen
  \bibfield  {author} {\bibinfo {author} {\bibfnamefont {D.~J.}\ \bibnamefont {Jacobs}}\ and\ \bibinfo {author} {\bibfnamefont {M.~F.}\ \bibnamefont {Thorpe}},\ }\href {https://doi.org/10.1103/PhysRevLett.75.4051} {\bibfield  {journal} {\bibinfo  {journal} {Physical Review Letters}\ }\textbf {\bibinfo {volume} {75}},\ \bibinfo {pages} {4051} (\bibinfo {year} {1995})}\BibitemShut {NoStop}%
\bibitem [{\citenamefont {Jacobs}\ and\ \citenamefont {Hendrickson}(1997)}]{jacobs_algorithm_1997}%
  \BibitemOpen
  \bibfield  {author} {\bibinfo {author} {\bibfnamefont {D.~J.}\ \bibnamefont {Jacobs}}\ and\ \bibinfo {author} {\bibfnamefont {B.}~\bibnamefont {Hendrickson}},\ }\href {https://doi.org/10.1006/jcph.1997.5809} {\bibfield  {journal} {\bibinfo  {journal} {Journal of Computational Physics}\ }\textbf {\bibinfo {volume} {137}},\ \bibinfo {pages} {346} (\bibinfo {year} {1997})}\BibitemShut {NoStop}%
\bibitem [{\citenamefont {Henkes}\ \emph {et~al.}(2016)\citenamefont {Henkes}, \citenamefont {Quint}, \citenamefont {Fily},\ and\ \citenamefont {Schwarz}}]{henkes_rigid_2016}%
  \BibitemOpen
  \bibfield  {author} {\bibinfo {author} {\bibfnamefont {S.}~\bibnamefont {Henkes}}, \bibinfo {author} {\bibfnamefont {D.~A.}\ \bibnamefont {Quint}}, \bibinfo {author} {\bibfnamefont {Y.}~\bibnamefont {Fily}},\ and\ \bibinfo {author} {\bibfnamefont {J.}~\bibnamefont {Schwarz}},\ }\href {https://doi.org/10.1103/PhysRevLett.116.028301} {\bibfield  {journal} {\bibinfo  {journal} {Physical Review Letters}\ }\textbf {\bibinfo {volume} {116}},\ \bibinfo {pages} {028301} (\bibinfo {year} {2016})}\BibitemShut {NoStop}%
\bibitem [{\citenamefont {Liu}\ \emph {et~al.}(2019)\citenamefont {Liu}, \citenamefont {Henkes},\ and\ \citenamefont {Schwarz}}]{liu_frictional_2019}%
  \BibitemOpen
  \bibfield  {author} {\bibinfo {author} {\bibfnamefont {K.}~\bibnamefont {Liu}}, \bibinfo {author} {\bibfnamefont {S.}~\bibnamefont {Henkes}},\ and\ \bibinfo {author} {\bibfnamefont {J.~M.}\ \bibnamefont {Schwarz}},\ }\href {https://doi.org/10.1103/PhysRevX.9.021006} {\bibfield  {journal} {\bibinfo  {journal} {Physical Review X}\ }\textbf {\bibinfo {volume} {9}},\ \bibinfo {pages} {021006} (\bibinfo {year} {2019})}\BibitemShut {NoStop}%
\bibitem [{\citenamefont {van~der Naald}\ \emph {et~al.}(2024)\citenamefont {van~der Naald}, \citenamefont {Singh}, \citenamefont {Eid}, \citenamefont {Tang}, \citenamefont {de~Pablo},\ and\ \citenamefont {Jaeger}}]{van_der_naald_minimally_2024}%
  \BibitemOpen
  \bibfield  {author} {\bibinfo {author} {\bibfnamefont {M.}~\bibnamefont {van~der Naald}}, \bibinfo {author} {\bibfnamefont {A.}~\bibnamefont {Singh}}, \bibinfo {author} {\bibfnamefont {T.~T.}\ \bibnamefont {Eid}}, \bibinfo {author} {\bibfnamefont {K.}~\bibnamefont {Tang}}, \bibinfo {author} {\bibfnamefont {J.~J.}\ \bibnamefont {de~Pablo}},\ and\ \bibinfo {author} {\bibfnamefont {H.~M.}\ \bibnamefont {Jaeger}},\ }\href {https://doi.org/10.1038/s41567-023-02354-3} {\bibfield  {journal} {\bibinfo  {journal} {Nature Physics}\ }\textbf {\bibinfo {volume} {20}},\ \bibinfo {pages} {653} (\bibinfo {year} {2024})}\BibitemShut {NoStop}%
\bibitem [{\citenamefont {Lester}\ and\ \citenamefont {Li}(2018)}]{lester_frictional_2018}%
  \BibitemOpen
  \bibfield  {author} {\bibinfo {author} {\bibfnamefont {D.}~\bibnamefont {Lester}}\ and\ \bibinfo {author} {\bibfnamefont {R.}~\bibnamefont {Li}},\ }\href {https://doi.org/10.1016/j.jcp.2018.05.016} {\bibfield  {journal} {\bibinfo  {journal} {Journal of Computational Physics}\ }\textbf {\bibinfo {volume} {369}},\ \bibinfo {pages} {225} (\bibinfo {year} {2018})}\BibitemShut {NoStop}%
\bibitem [{\citenamefont {Chubynsky}\ and\ \citenamefont {Thorpe}(2007)}]{chubynsky_algorithms_2007}%
  \BibitemOpen
  \bibfield  {author} {\bibinfo {author} {\bibfnamefont {M.~V.}\ \bibnamefont {Chubynsky}}\ and\ \bibinfo {author} {\bibfnamefont {M.~F.}\ \bibnamefont {Thorpe}},\ }\href {https://doi.org/10.1103/PhysRevE.76.041135} {\bibfield  {journal} {\bibinfo  {journal} {Physical Review E}\ }\textbf {\bibinfo {volume} {76}},\ \bibinfo {pages} {041135} (\bibinfo {year} {2007})}\BibitemShut {NoStop}%
\end{thebibliography}%


\begin{thebibliography}{16}%
\makeatletter
\providecommand \@ifxundefined [1]{%
 \@ifx{#1\undefined}
}%
\providecommand \@ifnum [1]{%
 \ifnum #1\expandafter \@firstoftwo
 \else \expandafter \@secondoftwo
 \fi
}%
\providecommand \@ifx [1]{%
 \ifx #1\expandafter \@firstoftwo
 \else \expandafter \@secondoftwo
 \fi
}%
\providecommand \natexlab [1]{#1}%
\providecommand \enquote  [1]{``#1''}%
\providecommand \bibnamefont  [1]{#1}%
\providecommand \bibfnamefont [1]{#1}%
\providecommand \citenamefont [1]{#1}%
\providecommand \href@noop [0]{\@secondoftwo}%
\providecommand \href [0]{\begingroup \@sanitize@url \@href}%
\providecommand \@href[1]{\@@startlink{#1}\@@href}%
\providecommand \@@href[1]{\endgroup#1\@@endlink}%
\providecommand \@sanitize@url [0]{\catcode `\\12\catcode `\$12\catcode `\&12\catcode `\#12\catcode `\^12\catcode `\_12\catcode `\%12\relax}%
\providecommand \@@startlink[1]{}%
\providecommand \@@endlink[0]{}%
\providecommand \url  [0]{\begingroup\@sanitize@url \@url }%
\providecommand \@url [1]{\endgroup\@href {#1}{\urlprefix }}%
\providecommand \urlprefix  [0]{URL }%
\providecommand \Eprint [0]{\href }%
\providecommand \doibase [0]{https://doi.org/}%
\providecommand \selectlanguage [0]{\@gobble}%
\providecommand \bibinfo  [0]{\@secondoftwo}%
\providecommand \bibfield  [0]{\@secondoftwo}%
\providecommand \translation [1]{[#1]}%
\providecommand \BibitemOpen [0]{}%
\providecommand \bibitemStop [0]{}%
\providecommand \bibitemNoStop [0]{.\EOS\space}%
\providecommand \EOS [0]{\spacefactor3000\relax}%
\providecommand \BibitemShut  [1]{\csname bibitem#1\endcsname}%
\let\auto@bib@innerbib\@empty
\bibitem [{\citenamefont {Kool}\ \emph {et~al.}(2025)\citenamefont {Kool}, \citenamefont {Tampier}, \citenamefont {Bourrianne},\ and\ \citenamefont {Lindner}}]{kool_hopper_2025}%
  \BibitemOpen
  \bibfield  {author} {\bibinfo {author} {\bibfnamefont {L.}~\bibnamefont {Kool}}, \bibinfo {author} {\bibfnamefont {J.}~\bibnamefont {Tampier}}, \bibinfo {author} {\bibfnamefont {P.}~\bibnamefont {Bourrianne}},\ and\ \bibinfo {author} {\bibfnamefont {A.}~\bibnamefont {Lindner}},\ }\href {https://doi.org/10.1017/flo.2025.10019} {\bibfield  {journal} {\bibinfo  {journal} {Flow}\ }\textbf {\bibinfo {volume} {5}},\ \bibinfo {pages} {E26} (\bibinfo {year} {2025})}\BibitemShut {NoStop}%
\bibitem [{\citenamefont {Unger}\ \emph {et~al.}(2000)\citenamefont {Unger}, \citenamefont {Chou}, \citenamefont {Thorsen}, \citenamefont {Scherer},\ and\ \citenamefont {Quake}}]{unger_monolithic_2000}%
  \BibitemOpen
  \bibfield  {author} {\bibinfo {author} {\bibfnamefont {M.~A.}\ \bibnamefont {Unger}}, \bibinfo {author} {\bibfnamefont {H.~P.}\ \bibnamefont {Chou}}, \bibinfo {author} {\bibfnamefont {T.}~\bibnamefont {Thorsen}}, \bibinfo {author} {\bibfnamefont {A.}~\bibnamefont {Scherer}},\ and\ \bibinfo {author} {\bibfnamefont {S.~R.}\ \bibnamefont {Quake}},\ }\href {https://doi.org/10.1126/science.288.5463.113} {\bibfield  {journal} {\bibinfo  {journal} {Science (New York, N.Y.)}\ }\textbf {\bibinfo {volume} {288}},\ \bibinfo {pages} {113} (\bibinfo {year} {2000})}\BibitemShut {NoStop}%
\bibitem [{\citenamefont {Dendukuri}\ \emph {et~al.}(2007)\citenamefont {Dendukuri}, \citenamefont {Gu}, \citenamefont {Pregibon}, \citenamefont {Hatton},\ and\ \citenamefont {Doyle}}]{dendukuri_stop-flow_2007}%
  \BibitemOpen
  \bibfield  {author} {\bibinfo {author} {\bibfnamefont {D.}~\bibnamefont {Dendukuri}}, \bibinfo {author} {\bibfnamefont {S.~S.}\ \bibnamefont {Gu}}, \bibinfo {author} {\bibfnamefont {D.~C.}\ \bibnamefont {Pregibon}}, \bibinfo {author} {\bibfnamefont {T.~A.}\ \bibnamefont {Hatton}},\ and\ \bibinfo {author} {\bibfnamefont {P.~S.}\ \bibnamefont {Doyle}},\ }\href {https://doi.org/10.1039/B703457A} {\bibfield  {journal} {\bibinfo  {journal} {Lab on a Chip}\ }\textbf {\bibinfo {volume} {7}},\ \bibinfo {pages} {818} (\bibinfo {year} {2007})}\BibitemShut {NoStop}%
\bibitem [{\citenamefont {Cappello}\ \emph {et~al.}(2019)\citenamefont {Cappello}, \citenamefont {Bechert}, \citenamefont {Duprat}, \citenamefont {du~Roure}, \citenamefont {Gallaire},\ and\ \citenamefont {Lindner}}]{cappello_transport_2019}%
  \BibitemOpen
  \bibfield  {author} {\bibinfo {author} {\bibfnamefont {J.}~\bibnamefont {Cappello}}, \bibinfo {author} {\bibfnamefont {M.}~\bibnamefont {Bechert}}, \bibinfo {author} {\bibfnamefont {C.}~\bibnamefont {Duprat}}, \bibinfo {author} {\bibfnamefont {O.}~\bibnamefont {du~Roure}}, \bibinfo {author} {\bibfnamefont {F.}~\bibnamefont {Gallaire}},\ and\ \bibinfo {author} {\bibfnamefont {A.}~\bibnamefont {Lindner}},\ }\href {https://doi.org/10.1103/PhysRevFluids.4.034202} {\bibfield  {journal} {\bibinfo  {journal} {Physical Review Fluids}\ }\textbf {\bibinfo {volume} {4}},\ \bibinfo {pages} {034202} (\bibinfo {year} {2019})}\BibitemShut {NoStop}%
\bibitem [{\citenamefont {Allan}\ \emph {et~al.}(2024)\citenamefont {Allan}, \citenamefont {Caswell}, \citenamefont {Keim}, \citenamefont {van~der Wel},\ and\ \citenamefont {Verweij}}]{allan_soft-mattertrackpy_2024}%
  \BibitemOpen
  \bibfield  {author} {\bibinfo {author} {\bibfnamefont {D.~B.}\ \bibnamefont {Allan}}, \bibinfo {author} {\bibfnamefont {T.}~\bibnamefont {Caswell}}, \bibinfo {author} {\bibfnamefont {N.~C.}\ \bibnamefont {Keim}}, \bibinfo {author} {\bibfnamefont {C.~M.}\ \bibnamefont {van~der Wel}},\ and\ \bibinfo {author} {\bibfnamefont {R.~W.}\ \bibnamefont {Verweij}},\ }\href {https://doi.org/10.5281/zenodo.12708864} {\bibinfo {title} {soft-matter/trackpy: v0.6.4}} (\bibinfo {year} {2024})\BibitemShut {NoStop}%
\bibitem [{\citenamefont {Lerner}\ \emph {et~al.}(2013)\citenamefont {Lerner}, \citenamefont {Düring},\ and\ \citenamefont {Wyart}}]{lerner_simulations_2013}%
  \BibitemOpen
  \bibfield  {author} {\bibinfo {author} {\bibfnamefont {E.}~\bibnamefont {Lerner}}, \bibinfo {author} {\bibfnamefont {G.}~\bibnamefont {Düring}},\ and\ \bibinfo {author} {\bibfnamefont {M.}~\bibnamefont {Wyart}},\ }\href {https://doi.org/10.1016/j.cpc.2012.10.020} {\bibfield  {journal} {\bibinfo  {journal} {Computer Physics Communications}\ }\textbf {\bibinfo {volume} {184}},\ \bibinfo {pages} {628} (\bibinfo {year} {2013})}\BibitemShut {NoStop}%
\bibitem [{\citenamefont {Thompson}\ \emph {et~al.}(2022)\citenamefont {Thompson}, \citenamefont {Aktulga}, \citenamefont {Berger}, \citenamefont {Bolintineanu}, \citenamefont {Brown}, \citenamefont {Crozier}, \citenamefont {in~'t Veld}, \citenamefont {Kohlmeyer}, \citenamefont {Moore}, \citenamefont {Nguyen}, \citenamefont {Shan}, \citenamefont {Stevens}, \citenamefont {Tranchida}, \citenamefont {Trott},\ and\ \citenamefont {Plimpton}}]{thompson_lammps_2022}%
  \BibitemOpen
  \bibfield  {author} {\bibinfo {author} {\bibfnamefont {A.~P.}\ \bibnamefont {Thompson}}, \bibinfo {author} {\bibfnamefont {H.~M.}\ \bibnamefont {Aktulga}}, \bibinfo {author} {\bibfnamefont {R.}~\bibnamefont {Berger}}, \bibinfo {author} {\bibfnamefont {D.~S.}\ \bibnamefont {Bolintineanu}}, \bibinfo {author} {\bibfnamefont {W.~M.}\ \bibnamefont {Brown}}, \bibinfo {author} {\bibfnamefont {P.~S.}\ \bibnamefont {Crozier}}, \bibinfo {author} {\bibfnamefont {P.~J.}\ \bibnamefont {in~'t Veld}}, \bibinfo {author} {\bibfnamefont {A.}~\bibnamefont {Kohlmeyer}}, \bibinfo {author} {\bibfnamefont {S.~G.}\ \bibnamefont {Moore}}, \bibinfo {author} {\bibfnamefont {T.~D.}\ \bibnamefont {Nguyen}}, \bibinfo {author} {\bibfnamefont {R.}~\bibnamefont {Shan}}, \bibinfo {author} {\bibfnamefont {M.~J.}\ \bibnamefont {Stevens}}, \bibinfo {author} {\bibfnamefont {J.}~\bibnamefont {Tranchida}}, \bibinfo {author} {\bibfnamefont {C.}~\bibnamefont {Trott}},\ and\ \bibinfo {author} {\bibfnamefont {S.~J.}\ \bibnamefont {Plimpton}},\ }\href
  {https://doi.org/10.1016/j.cpc.2021.108171} {\bibfield  {journal} {\bibinfo  {journal} {Computer Physics Communications}\ }\textbf {\bibinfo {volume} {271}},\ \bibinfo {pages} {108171} (\bibinfo {year} {2022})}\BibitemShut {NoStop}%
\bibitem [{\citenamefont {Luding}(2008)}]{luding_cohesive_2008}%
  \BibitemOpen
  \bibfield  {author} {\bibinfo {author} {\bibfnamefont {S.}~\bibnamefont {Luding}},\ }\href {https://doi.org/10.1007/s10035-008-0099-x} {\bibfield  {journal} {\bibinfo  {journal} {Granular Matter}\ }\textbf {\bibinfo {volume} {10}},\ \bibinfo {pages} {235} (\bibinfo {year} {2008})}\BibitemShut {NoStop}%
\bibitem [{\citenamefont {Ai}\ \emph {et~al.}(2011)\citenamefont {Ai}, \citenamefont {Chen}, \citenamefont {Rotter},\ and\ \citenamefont {Ooi}}]{ai_assessment_2011}%
  \BibitemOpen
  \bibfield  {author} {\bibinfo {author} {\bibfnamefont {J.}~\bibnamefont {Ai}}, \bibinfo {author} {\bibfnamefont {J.-F.}\ \bibnamefont {Chen}}, \bibinfo {author} {\bibfnamefont {J.~M.}\ \bibnamefont {Rotter}},\ and\ \bibinfo {author} {\bibfnamefont {J.~Y.}\ \bibnamefont {Ooi}},\ }\href {https://doi.org/10.1016/j.powtec.2010.09.030} {\bibfield  {journal} {\bibinfo  {journal} {Powder Technology}\ }\textbf {\bibinfo {volume} {206}},\ \bibinfo {pages} {269} (\bibinfo {year} {2011})}\BibitemShut {NoStop}%
\bibitem [{\citenamefont {Shäfer}\ \emph {et~al.}(1996)\citenamefont {Shäfer}, \citenamefont {Dippel},\ and\ \citenamefont {Wolf}}]{shafer_force_1996}%
  \BibitemOpen
  \bibfield  {author} {\bibinfo {author} {\bibfnamefont {J.}~\bibnamefont {Shäfer}}, \bibinfo {author} {\bibfnamefont {S.}~\bibnamefont {Dippel}},\ and\ \bibinfo {author} {\bibfnamefont {D.~E.}\ \bibnamefont {Wolf}},\ }\href {https://doi.org/10.1051/jp1:1996129} {\bibfield  {journal} {\bibinfo  {journal} {Journal de Physique I}\ }\textbf {\bibinfo {volume} {6}},\ \bibinfo {pages} {5} (\bibinfo {year} {1996})}\BibitemShut {NoStop}%
\bibitem [{\citenamefont {Jacobs}\ and\ \citenamefont {Thorpe}(1995)}]{jacobs_generic_1995}%
  \BibitemOpen
  \bibfield  {author} {\bibinfo {author} {\bibfnamefont {D.~J.}\ \bibnamefont {Jacobs}}\ and\ \bibinfo {author} {\bibfnamefont {M.~F.}\ \bibnamefont {Thorpe}},\ }\href {https://doi.org/10.1103/PhysRevLett.75.4051} {\bibfield  {journal} {\bibinfo  {journal} {Physical Review Letters}\ }\textbf {\bibinfo {volume} {75}},\ \bibinfo {pages} {4051} (\bibinfo {year} {1995})}\BibitemShut {NoStop}%
\bibitem [{\citenamefont {Jacobs}\ and\ \citenamefont {Hendrickson}(1997)}]{jacobs_algorithm_1997}%
  \BibitemOpen
  \bibfield  {author} {\bibinfo {author} {\bibfnamefont {D.~J.}\ \bibnamefont {Jacobs}}\ and\ \bibinfo {author} {\bibfnamefont {B.}~\bibnamefont {Hendrickson}},\ }\href {https://doi.org/10.1006/jcph.1997.5809} {\bibfield  {journal} {\bibinfo  {journal} {Journal of Computational Physics}\ }\textbf {\bibinfo {volume} {137}},\ \bibinfo {pages} {346} (\bibinfo {year} {1997})}\BibitemShut {NoStop}%
\bibitem [{\citenamefont {Henkes}\ \emph {et~al.}(2016)\citenamefont {Henkes}, \citenamefont {Quint}, \citenamefont {Fily},\ and\ \citenamefont {Schwarz}}]{henkes_rigid_2016}%
  \BibitemOpen
  \bibfield  {author} {\bibinfo {author} {\bibfnamefont {S.}~\bibnamefont {Henkes}}, \bibinfo {author} {\bibfnamefont {D.~A.}\ \bibnamefont {Quint}}, \bibinfo {author} {\bibfnamefont {Y.}~\bibnamefont {Fily}},\ and\ \bibinfo {author} {\bibfnamefont {J.}~\bibnamefont {Schwarz}},\ }\href {https://doi.org/10.1103/PhysRevLett.116.028301} {\bibfield  {journal} {\bibinfo  {journal} {Physical Review Letters}\ }\textbf {\bibinfo {volume} {116}},\ \bibinfo {pages} {028301} (\bibinfo {year} {2016})}\BibitemShut {NoStop}%
\bibitem [{\citenamefont {Henkes}(2026)}]{henkes_silkehenkesrigidlibrary_2026}%
  \BibitemOpen
  \bibfield  {author} {\bibinfo {author} {\bibfnamefont {S.}~\bibnamefont {Henkes}},\ }\href {https://github.com/silkehenkes/RigidLibrary} {\bibinfo {title} {silkehenkes/{RigidLibrary}}} (\bibinfo {year} {2026}),\ \bibinfo {note} {original-date: 2018-08-29T13:03:16Z}\BibitemShut {NoStop}%
\bibitem [{\citenamefont {Wyart}\ \emph {et~al.}(2005)\citenamefont {Wyart}, \citenamefont {Nagel},\ and\ \citenamefont {Witten}}]{wyart_geometric_2005}%
  \BibitemOpen
  \bibfield  {author} {\bibinfo {author} {\bibfnamefont {M.}~\bibnamefont {Wyart}}, \bibinfo {author} {\bibfnamefont {S.~R.}\ \bibnamefont {Nagel}},\ and\ \bibinfo {author} {\bibfnamefont {T.~A.}\ \bibnamefont {Witten}},\ }\href {https://doi.org/10.1209/epl/i2005-10245-5} {\bibfield  {journal} {\bibinfo  {journal} {Europhysics Letters}\ }\textbf {\bibinfo {volume} {72}},\ \bibinfo {pages} {486} (\bibinfo {year} {2005})}\BibitemShut {NoStop}%
\bibitem [{\citenamefont {Goodrich}\ \emph {et~al.}(2013)\citenamefont {Goodrich}, \citenamefont {Ellenbroek},\ and\ \citenamefont {Liu}}]{goodrich_stability_2013}%
  \BibitemOpen
  \bibfield  {author} {\bibinfo {author} {\bibfnamefont {C.~P.}\ \bibnamefont {Goodrich}}, \bibinfo {author} {\bibfnamefont {W.~G.}\ \bibnamefont {Ellenbroek}},\ and\ \bibinfo {author} {\bibfnamefont {A.~J.}\ \bibnamefont {Liu}},\ }\href {https://doi.org/10.1039/C3SM51095F} {\bibfield  {journal} {\bibinfo  {journal} {Soft Matter}\ }\textbf {\bibinfo {volume} {9}},\ \bibinfo {pages} {10993} (\bibinfo {year} {2013})}\BibitemShut {NoStop}%
\end{thebibliography}%


\begin{thebibliography}{2}%
\makeatletter
\providecommand \@ifxundefined [1]{%
 \@ifx{#1\undefined}
}%
\providecommand \@ifnum [1]{%
 \ifnum #1\expandafter \@firstoftwo
 \else \expandafter \@secondoftwo
 \fi
}%
\providecommand \@ifx [1]{%
 \ifx #1\expandafter \@firstoftwo
 \else \expandafter \@secondoftwo
 \fi
}%
\providecommand \natexlab [1]{#1}%
\providecommand \enquote  [1]{``#1''}%
\providecommand \bibnamefont  [1]{#1}%
\providecommand \bibfnamefont [1]{#1}%
\providecommand \citenamefont [1]{#1}%
\providecommand \href@noop [0]{\@secondoftwo}%
\providecommand \href [0]{\begingroup \@sanitize@url \@href}%
\providecommand \@href[1]{\@@startlink{#1}\@@href}%
\providecommand \@@href[1]{\endgroup#1\@@endlink}%
\providecommand \@sanitize@url [0]{\catcode `\\12\catcode `\$12\catcode `\&12\catcode `\#12\catcode `\^12\catcode `\_12\catcode `\%12\relax}%
\providecommand \@@startlink[1]{}%
\providecommand \@@endlink[0]{}%
\providecommand \url  [0]{\begingroup\@sanitize@url \@url }%
\providecommand \@url [1]{\endgroup\@href {#1}{\urlprefix }}%
\providecommand \urlprefix  [0]{URL }%
\providecommand \Eprint [0]{\href }%
\providecommand \doibase [0]{https://doi.org/}%
\providecommand \selectlanguage [0]{\@gobble}%
\providecommand \bibinfo  [0]{\@secondoftwo}%
\providecommand \bibfield  [0]{\@secondoftwo}%
\providecommand \translation [1]{[#1]}%
\providecommand \BibitemOpen [0]{}%
\providecommand \bibitemStop [0]{}%
\providecommand \bibitemNoStop [0]{.\EOS\space}%
\providecommand \EOS [0]{\spacefactor3000\relax}%
\providecommand \BibitemShut  [1]{\csname bibitem#1\endcsname}%
\let\auto@bib@innerbib\@empty
\bibitem [{\citenamefont {To}\ \emph {et~al.}(2001)\citenamefont {To}, \citenamefont {Lai},\ and\ \citenamefont {Pak}}]{to_jamming_2001}%
  \BibitemOpen
  \bibfield  {author} {\bibinfo {author} {\bibfnamefont {K.}~\bibnamefont {To}}, \bibinfo {author} {\bibfnamefont {P.-Y.}\ \bibnamefont {Lai}},\ and\ \bibinfo {author} {\bibfnamefont {H.~K.}\ \bibnamefont {Pak}},\ }\href {https://doi.org/10.1103/PhysRevLett.86.71} {\bibfield  {journal} {\bibinfo  {journal} {Physical Review Letters}\ }\textbf {\bibinfo {volume} {86}},\ \bibinfo {pages} {71} (\bibinfo {year} {2001})}\BibitemShut {NoStop}%
\bibitem [{\citenamefont {Garcimartín}\ \emph {et~al.}(2010)\citenamefont {Garcimartín}, \citenamefont {Zuriguel}, \citenamefont {Pugnaloni},\ and\ \citenamefont {Janda}}]{garcimartin_shape_2010}%
  \BibitemOpen
  \bibfield  {author} {\bibinfo {author} {\bibfnamefont {A.}~\bibnamefont {Garcimartín}}, \bibinfo {author} {\bibfnamefont {I.}~\bibnamefont {Zuriguel}}, \bibinfo {author} {\bibfnamefont {L.~A.}\ \bibnamefont {Pugnaloni}},\ and\ \bibinfo {author} {\bibfnamefont {A.}~\bibnamefont {Janda}},\ }\href {https://doi.org/10.1103/PhysRevE.82.031306} {\bibfield  {journal} {\bibinfo  {journal} {Physical Review E}\ }\textbf {\bibinfo {volume} {82}},\ \bibinfo {pages} {031306} (\bibinfo {year} {2010})}\BibitemShut {NoStop}%
\end{thebibliography}%
\end{document}